\documentclass[12pt]{article}
\usepackage{amssymb}
\begin{document}

\def\squeeze{\hspace{-1.5em}}
\def\bra#1{\langle #1 \vert\,}
\def\ket#1{\,\vert #1 \rangle}
\def\vev#1{\langle #1 \rangle}
\def\st{\, : \,}
\def\kbar{{\mathchar'26\mkern-9muk}}  
\def\tr{\mbox{Tr}}
\def\ad{\mbox{ad}\,}
\def\ker{\mbox{Ker}\,}
\def\b#1{{\mathbb #1}}
\def\c#1{{\cal #1}}

\newcommand{\sect}[1]{\setcounter{equation}{0}\section{#1}}
\renewcommand{\theequation}{\thesubsection.\arabic{equation}}
\renewcommand{\theequation}{\thesection.\arabic{equation}}

\newcommand{\be}{\begin{equation}}
\newcommand{\ee}{\end{equation}}
\newcommand{\bea}{\begin{eqnarray}}
\newcommand{\eea}{\end{eqnarray}}

\title{Finite Field Theory on Noncommutative Geometries}

\author{S. Cho$\strut^1$ \, R. Hinterding$\strut^{2,3}$ \, 
        J. Madore$\strut^{3,4}$ \, H. Steinacker$\strut^2$
        \and
        $\strut^1$Department of Physics, Semyung University \\
        Chechon, Chungbuk 390 - 711, Korea
        \and
        $\strut^2$Sektion Physik, Ludwig-Maximilian Universit\"at\\
        Theresienstra\ss e 37, D-80333 M\"unchen
        \and
        $\strut^3$Max-Planck-Institut f\"ur Physik\\
        (Werner-Heisenberg-Institut)\\
        F\"ohringer Ring 6, D-80805 M\"unchen
        \and
        $\strut^4$Laboratoire de Physique Th\'eorique et Hautes Energies\\
        Universit\'e de Paris-Sud, B\^atiment 211, F-91405 Orsay
        }

\maketitle

\begin{abstract}
The propagator is calculated on a noncommutative version of the
flat plane and the Lobachevsky plane with and without an extra
(euclidean) time parameter. In agreement with the general idea of
noncommutative geometry it is found that the limit when the two
`points' coincide is finite and diverges only when the geometry
becomes commutative.  The flat 4-dimensional case is also
considered. This is at the moment less interesting since there has
been no curved case developed with which it can be compared.
\end{abstract}

\parskip 4pt plus2pt minus2pt
\vfill
\noindent
LMU-TPW 99-06\\
\noindent
MPI-PhT/99-12
\newpage

\sect{Introduction and motivation}

It was postulated some time ago~\cite{Sny47a, Sny47b} that a
noncommutative structure at small length scales could introduce an
effective cut-off in field theory similar to a lattice but at the same
time maintain Lorentz invariance. Recently there has been a revival of
this idea and several new examples~\cite{Mad92, DopFreRob95,
KehMeeZou95, AzcKulRod97, KosLukMas99, KraWul99} have been
studied. Models~\cite{FicLorWes96, KemManMan95, KemMan97,
CerHinMadWes99} in `1-dimension' have also added to our understanding
of the `lattice' structure.  The basic idea is simple and can be
illustrated by a classical particle moving in a plane, described by
two position coordinates $(q^1, q^2)$ and two momentum coordinates
$(p_1, p_2)$.  In the language of quantum mechanics these four
classical coordinates are commuting operators.  In the presence of a
magnetic field $B$ normal to the plane the momentum operators are
modified and they cease to commute:
\be
[p_1, p_2] = i \hbar eB.                                         \label{1.5}
\ee
This introduces a cellular structure in the momentum plane. It becomes
divided into Landau cells of area proportional to $\hbar eB$. Consider
in this case the divergent integral
$$
I = \int {dp_1 dp_2 \over p^2}.
$$
The commutation relation (\ref{1.5}) does not permit $p_1$ and $p_2$
simultaneously to take the eigenvalue zero and the operator 
$p^2 = p_1^2 + p_2^2$ is bounded below by $\hbar eB$. The magnetic field
acts as an infrared cut-off.  If the position space were curved, with
constant Gaussian curvature $K$ one would obtain again an infrared
regularization for $I$.  In an exactly analogous fashion, to obtain an
ultraviolet regularization one must replace the coordinates of position
space by two operators which do not commute:
\be
[q^1, q^2] = i\kbar q^{12}.                                       \label{1.7}
\ee
By the new uncertainty relation there is no longer a notion of a point
in position space since one cannot measure both coordinates
simultaneously but as before, position space can be thought of as
divided into Planck cells. It has become fuzzy.  This cellular structure
serves as an ultraviolet cut-off similar to a lattice structure. If we
consider for example the divergent integral $I$ and introduce also a
Gaussian curvature we find
\be
I \sim \log (\kbar K).                                           \label{1.8}
\ee
The integral has become completely regularized.  There is however now a
new complication; the right-hand side of (\ref{1.8}) seems not to depend
on the operator $q^{12}$. We have argued elsewhere~\cite{Mad99} that,
endowed with an appropriate differential structure, each fuzzy
space-time supports a uniquely determined gravitational field and that
the latter is a classical manifestation of the commutation relations
plus a differential structure. From this point of view what we put on
the right-hand side of (\ref{1.7}) will depend on which gravitational
field we wish to regularize the integral with. That is, in fact $K$ does
depend on $q^{12}$.  

In Section~2 we shall give a description of how the integral $I$ of
(\ref{1.8}) is to be calculated in the case of a general algebra
$\c{A}$. The propagator is an element of the tensor product 
$\c{H} \otimes \c{H}$ of two copies of a Hilbert space 
$\c{H} \subset \c{A}$. We represent $\c{A} \otimes \c{A}$ as an
algebra of operators on the tensor product 
$L^2(V,d\mu) \otimes L^2(V,d\mu)$ of two copies of another Hilbert
space $L^2(V,d\mu)$ of functions on a manifold $V$, square integrable
with respect to some measure $d\mu$. We then express 
$L^2(V,d\mu) \otimes L^2(V,d\mu)$ as the tensor product of a Hilbert
space $\c{D} \simeq L^2(V,d\mu)$, which represents the diagonal
elements of $\c{A} \otimes \c{A}$, and an extra Hilbert space $\c{F}$,
which describes the off-diagonal expansion. This must be done in a way
consistent with the commutation relations. Those of $\c{F}$
effectively force the distance from the diagonal in the tensor product
to be `quantized' and exclude the value zero. In the examples we shall
see that if one were to interpret a given set of matrix elements of
the propagator of the tensor product as a propagator on an ordinary
space then it would appear to be associated to a non-local
differential operator~\cite{Yuk49, PaiUhl50}. In Section~3 we apply
the formalism to the case of a noncommutative version~\cite{MadMou98}
of $\b{R}^2$ with a flat metric obtained by setting $q^{12} = 1$. In
Section~4 we shall be interested in a noncommutative
version~\cite{Lei96, ChoMadPar98} of the Lobachevsky half-plane, the
surface of constant negative Gaussian curvature. Finally in Section~5
we examine briefly the extension to dimension 4 and the problem of
Lorentz invariance. In this paper we consider infinite-dimensional
algebras. There are also models which are described by
finite-dimensional algebras~\cite{Mad99, Ste97} where the fact that
the $n$-point elements are well-defined is automatic.

\sect{The general theory}

In general consider any $*$-algebra $\c{A}$ with a trivial center, in
some representation with a partial trace and let $\Delta$ be a linear
operator on $\c{A}$ with a set of eigenvectors $\phi_r \in \c{A}$ and
corresponding real eigenvalues $\lambda_r$:
$$
\Delta \phi_r = \lambda_r \phi_r. 
$$
The parameter $r$ here designates a point in some parameter space and we
write the integral on this space as a sum over $r$.  The corresponding
classical action is
\be
S = \tr (\phi^* \Delta \phi), \qquad \phi \in \c{A}.        \label{action}
\ee
The trace here must be defined in some representation of $\c{A}$.  We 
shall assume that with respect to this trace
\be
\tr (\phi^*_r \phi^{\phantom{*}}_s) = \delta_{rs}          \label{ortho}
\ee
and we define the Hilbert space $\c{H} \subset \c{A}$ of 1-particle 
states to be
$$
\c{H} = \{ \phi = \sum_r a_r \phi_r \st \sum_r |a_r|^2 < \infty \}.
$$
As usual the $a_r$ become operators when the field is quantized.
For $f \in \c{H}$ the completeness condition can be written as
$$
\phi = \sum_r \phi_r \tr(\phi^*_r \phi).
$$
If we introduce the element 
$$
W = \sum_r \phi_r \otimes \phi^*_r
$$
then the completeness condition can also be written
$$
\tr_2 (W \cdot 1 \otimes \phi) = \phi \otimes 1.
$$
The tensor product is here over the complex numbers and the subscript 
on the trace indicates that it is taken over the second factor.  The 
element $W$ is therefore the noncommutative generalization of the
Dirac distribution in the commutative case; it is not an element of
$\c{H} \otimes \c{H}$. We introduce also the element $G$ defined by
the formal sum
\be
G = \sum \lambda_r^{-1}\phi_r \otimes \phi^*_r.                \label{defG}
\ee
Since obviously $\Delta G = W$ this element generalizes the propagator
corresponding to $\Delta$. We wish to discuss the conditions under
which the sum converges and $G$ can be considered as a
well-defined element of a weak closure of $\c{H} \otimes \c{H}$.

It is possible to give a second formal definition of $G$ using the
noncommutative version of the euclidean path integral. Let 
$S[\phi,J] = S[\phi] + \tr (J \phi)$ be the classical action of an
interacting scalar field in the presence of an external source 
$J \in \c{A}$. The term $S[\phi]$ would be a sum of the kinematical term
(\ref{action}) and an interaction term $S_J[\phi] = \tr(V(\phi))$ with
$V(\phi) \in \c{A}$.  Define the partition function $Z[J]$ and
generating functional $W[J]$ by
$$
Z[J] = \int d\phi e^{-S[\phi,J]} = e^{-W[J]}.
$$
If the algebra is for example a finite matrix algebra then this integral
can be considered as well defined. Otherwise we consider it as a
mnemonic trick. The theory is to be defined by the Gell-Mann-Low
expansion of the $n$-point elements in terms of the propagator,
with or without normal ordering.  The $n$-point element $G_{(n)}$ is
defined to be the functional derivative of $W[J]$ with respect to $J$:
$$
G_{(n)} = - {\delta^n W[J]\over \delta J_1 \cdots \delta J_n}.
$$
Here the $J_i$ are different occurrences of $J$. They are all canonically
equal to $J$ but carry an extra index to distinguish them:
$J_i = 1 \otimes \cdots \otimes J \otimes \cdots \otimes 1$
is an element of the $n$-fold tensor product of $\c{H}$.  By
construction $G_{(n)}$ also is an element of the $n$-fold tensor product
of $\c{H}$. In particular we have
$$
\vev{\phi}_J = Z[J]^{-1}\int d\phi \,\phi\, e^{-S[\phi,J]} =
- Z[J]^{-1} {\delta Z[J] \over \delta J} = 
{\delta W[J] \over \delta J}
$$
and
$$
\vev{\phi\otimes \phi}_J = 
Z[J]^{-1}\int d\phi \,\phi \otimes \phi \,e^{-S[\phi,J]} =
- {\delta W[J] \over \delta J_1\delta J_2}.
$$
If $S[\phi,J]$ is the free action then $\vev{\phi \otimes \phi}_0$ is
equal to the (bare) propagator $G$.  The bracket is here the quantum
bracket, which we distinguish with the index $J$. The context will
indicate whether $\phi$ designates a quantum operator or a classical
element of $\c{A}$.

With our definitions a composite field like $\phi^n \in \c{A}$ can
appear in the interaction term $S_I[\phi]$ of the action but
$\vev{\phi^n}_J$ is not defined. To define such objects we would, as in
the commutative case~\cite{Zin93}, introduce an extra source $J_{(n)}$
in the path integral and a corresponding extra term 
$\tr (\phi^n J_{(n)})$ in the action.  One might be tempted to define
for example $\vev{\phi^2}_J$ as the image of 
$\vev{\phi \otimes \phi}_J$ under the multiplication map 
$\pi:\,\c{A} \otimes \c{A} \to \c{A}$ but this will not be consistent in
the classical limit.  If one tries to define the expectation values of
composite fields in terms of $J$ one will come upon the same
divergences~\cite{Fil96, DopFreRob95, ChaDemPre98} as in ordinary field
theory. In the situations of interest the sum
$$
\pi G = \sum_r \lambda_r^{-1} |\phi_r|^2.
$$
diverges and it is not to be expected~\cite{VarGra98} that the
noncommutativity of the algebra will alter this fact.  We shall find
finite results because the noncommutativity `smears' the vertices, as
it does points in general.  By definition we have subtracted
disconnected `vacuum bubbles'. These could be singular; in the
commutative limit they would be proportional to the volume of
space-time. If the center of the algebra is not trivial one could 
still obtain a divergent result~\cite{KehMeeZou95}.

We shall restrict our attention to algebras which are generated by a set
$q^\mu$, $1 \leq \mu \leq n$, of $n$ hermitian elements. Define 
$q^{\mu\nu} \in \c{A}$ by
$$
[q^\mu, q^\nu] = i \kbar q^{\mu\nu}
$$
where $\kbar$ is a parameter which one can suppose to be of the order of
the square of the Planck length. This however is not necessary; the
experimental bounds are much weaker.  We shall suppose that $\c{A}$ is
represented as an algebra of operators on a Hilbert space $L^2(V,d\mu)$
and we fix an orthonormal basis $\ket{i}$. We can write then
$$
q^\mu \ket{i} =  \sum_j Q^\mu_{ji} \ket{j}
$$
for some set of $n$ matrices $Q^\mu_{ij}$. If the algebra is commutative
then $Q^\mu_{ij} = q^\mu_i \delta_{ij}$.  As above, the symbol $\Sigma$
here can represent a sum or an integral depending on the basis $\ket{i}$
it is convenient to choose. The index $i$ belongs again to some
parameter space which of course is not to be confused with the space to
which the parameters $r$ and $s$ of (\ref{ortho}) belong. The symbol 
$\delta_{ij}$ can represent therefore the Kronecker or Dirac delta. 

Consider the differential $d_u$ of the universal calculus. It is a map of
$\c{A}$ into $\c{A} \otimes \c{A}$ given by 
$d_u f = 1 \otimes f - f \otimes 1$. We define the `variation' 
$\delta q^\mu$ of the generator $q^\mu$ as
\be
\delta q^\mu = \frac 12 d_u q^\mu = 
\frac 12 (1 \otimes q^\mu - q^\mu \otimes 1).              \label{delta-q}
\ee
We identify $q^\mu = q^\mu \otimes 1$ in the tensor product and we
set $q^{\mu\prime} = 1 \otimes q^\mu$. Thus we can write
$$
\delta q^\mu = \frac 12 (q^{\mu\prime} - q^\mu).
$$
It follows from the commutation rules of the algebra that
$$
[\delta q^\mu, \delta q^\nu] = \frac 14 i \kbar  
(q^{\mu\nu} \otimes 1 + 1 \otimes q^{\mu\nu}).
$$
Suppose that a set of elements $\bar q^\mu$ of $\c{A} \otimes \c{A}$
can be found such that $\c{A} \otimes \c{A}$ is generated by the set
$\{\bar q^\mu, \delta q^\mu\}$ and such that
\be
[\bar q^\mu, \delta q^\nu] = 0.                               \label{b-d}
\ee
Then we can write the tensor product $L^2(V,d\mu) \otimes L^2(V,d\mu)$
in the form
\be
L^2(V,d\mu) \otimes L^2(V,d\mu) \simeq \c{D} \otimes \c{F} \label{factor}
\ee
where $\bar q^\mu$ acts on $\c{D}$ and $\delta q^\mu$ on $\c{F}$. We
shall choose accordingly a basis 
$$
\ket{\bar i,k} = \ket{\bar i}_D \otimes \ket{k}_F
$$
of $L^2(V,d\mu) \otimes L^2(V,d\mu)$. If $q^{\mu\nu}$ lies in the
center of the algebra then the elements
$$
\bar q^\mu = \frac 12 (q^\mu + q^{\mu\prime})
$$
are such that Equation~(\ref{b-d}) is satisfied. Further one has
$$
q^\mu = \bar q^\mu - \delta q^\mu, \qquad 
q^{\mu\prime} = \bar q^\mu + \delta q^\mu
$$
and with the obvious identifications
\be
[\bar q^\mu, \bar q^\nu] = \frac 12 i \kbar q^{\mu\nu},\qquad
[\delta q^\mu, \delta q^\nu] = \frac 12 i \kbar q^{\mu\nu}.  \label{diff-rel}
\ee
The tensor product in the definition of $G$ is now to
be considered as a tensor product of a `diagonal' algebra $\bar \c{A}$,
acting on $\c{D}$ and a `variation' $\delta \c{A}$, acting on $\c{F}$.
That is, we rewrite
\be
\c{A} \otimes \c{A} = \bar \c{A} \otimes \delta \c{A}      \label{bar-delta} 
\ee
in accordance with (\ref{factor}). If (\ref{b-d}) is not satisfied the
factorization (\ref{factor}) can still be of interest if $\delta q^\mu$
acts only on $\c{F}$. In general then $\bar q^\mu$ will act
non-trivially on the complete tensor product $\c{D} \otimes \c{F}$. We
shall suppose that the definition (\ref{delta-q}) of $\delta q^\mu$ in terms
of the tensor product coincides with the intuitive notion of the
`variation of a coordinate'.  One can introduce a new differential
calculus $(\bar\Omega^*(\c{A}), \bar d)$ defined by 
\be
\bar d \bar q^\mu = \delta q^\mu.                           \label{d-bar}
\ee
We shall see an example of this in Section~4. One would like this new
calculus to be isomorphic to the original one if $\delta q^\mu$ and 
$d q^\mu$ are to be thought of as `infinitesimal variations'.

Let $\c{C}(M)$ be an algebra of functions on a space $M$. Let $f$ be a
map of $M$ into itself and let $f^*$ be the induced map of $\c{C}(M)$
into itself. We set $\phi^\prime = f^*(\phi)$ and define 
$\delta \phi = \phi^\prime - \phi$. The ordinary propagator is a
function of two points, an element of $\c{C}(M) \otimes \c{C}(M)$ and we are
interested in the limit when the two points coincide. This limit must
be taken with care since the partial derivative of a function after
the limit and the limit of the derived function with respect to one of
the variables are not in general equal. We are interested in the
latter since the Laplace operator which defines the propagator
acts only on one of the variables. If we set $\delta x = x^\prime - x$
where $x^\prime = f(x)$ then we can express the limit $\delta x \to 0$
as $\delta \phi \to 0$.  We wish to study the element
$G(q^\mu;q^{\nu\prime})$ of the tensor product $\c{H} \otimes \c{H}$
most particularly in the limit $q^{\mu\prime} \to q^\mu$. The
$q^\mu$ are however fixed generators of the algebra and this limit must
be defined otherwise.  As a possible added complication, which will
however not appear explicitly in the examples we shall consider, the
generators $q^\mu$ are in general unbounded operators. We shall give a
formal definition of the limit as a weak limit within the tensor product
in terms of variations of the basis vectors $\ket{i}$. We shall
use a tensor product which is not braided. We shall return to his
assumption later.

Using the representation of $\c{A}$ the propagator 
$G = G(q^\mu;q^{\nu\prime})$ can be expressed as a map
$$
G:\;L^2(V,d\mu) \otimes L^2(V,d\mu)
\rightarrow L^2(V,d\mu) \otimes L^2(V,d\mu).
$$
It can be defined in terms of its (classical) matrix elements 
$\bra{j,j^\prime}G(q^\mu;q^{\nu\prime})\ket{i,i^\prime}$.  
In the commutative limit $\kbar\to 0$ one would find
$$
\bra{j,j^\prime} G(q^\mu;q^{\nu\prime})\ket{i,i^\prime}\to
G(q^\mu;q^{\nu\prime})\,\delta_{ij}\delta_{i^\prime j^\prime}
$$
with
$$
q^\mu\ket{i} = q^\mu_i\ket{i}, \qquad
q^{\nu\prime} \ket{i^\prime} = q^{\nu\prime}_{i^\prime} \ket{i^\prime}
$$
and so, at least in a quasicommutative approximation, we can identify
$q^\mu$ with a point $i \in V = \b{R}^n$ and $q^{\mu\prime}$ with
$i^\prime \in V = \b{R}^n$. We shall therefore represent graphically
$G(q^\mu; q^{\mu\prime})$ as a line between $i$ and $i^\prime$:
\be
\begin{array}{ccc}
\hbox to0pt{\hss$j$\hss}
&&\hbox to0pt{\hss$j^\prime$\hss}\\[-2pt]
\circ&\hskip -.7em
\vrule height .6ex width 3em depth -.45ex\hskip-.7em&
\circ\\[-2pt]
\hbox to0pt{\hss$i$\hss}
&&\hbox to0pt{\hss$i^\prime$\hss}
\end{array}                                              \label{Feynman1}
\ee 
The extra pair of indices $(j,j^\prime)$ is present because in general
$G$ acts as an operator on each end of the line.  An ordinary propagator
on a manifold diverges in the limit $q^{\mu\prime} \to q^\mu$. This
limit can be redefined as the limit
$$
\ket{i^\prime} \to \ket{i}.
$$
This limit makes sense in the noncommutative case but it cannot be
attained as we shall see below. We shall use therefore the
identification (\ref{factor}) to express the limit as
\be
\ket{\bar i,k} \to \ket{\bar i,0} \equiv \ket{\bar i}.        \label{join}
\ee
In the graph (\ref{Feynman1}) this means that the two ends of the line
almost close to form a circle. It does not really follow that $\ket{j}$
and $\ket{j^\prime}$ are related, except in the commutative limit. We
shall however suppose that
\be
\ket{\bar j,k} \to \ket{\bar j,0} \equiv \ket{\bar j} 
\ee
with (\ref{join}).

It is here that the representation, especially the representation of the
tensor product, becomes of importance. We shall describe the second copy
$\c{F}$ of the Hilbert space using creation and annihilation
operators. We choose then the basis $\ket{k}_F$ with $k\in \b{Z}_+$. The
states $\ket{\bar i,0}$ are those in which collectively the operators
$\delta q^\mu$ take their minimum value. If we introduce a distance $s$
by
$$
s^2 = g_{\mu\nu} \delta q^\mu \delta q^\nu
$$
then we can define the coincidence limit as a state in $\c{F}$ on which
$s$ takes its minimum value.  In the language of quantum mechanics such
a state is an example of a coherent state.

We introduce a set of $n$ annihilation operators $a_l$ with their adjoints
$a^*_m$ such that, as in quantum mechanics
\be
[a^{\phantom{*}}_l,a^*_m] = \kbar \delta_{lm}.                 \label{a-a*}
\ee
We shall see that each $a_l$ annihilates and each $a^*_l$ creates a unit
of separation.  The quantum mechanical analogue of this separation would
be the energy of the harmonic oscillator. By analogy then we define a
diagonal state to be a state annihilated by all the $a_l$. We define as
usual the action of $a_l$ on the diagonal basis element 
$\ket{\bar i,0} \in \c{D} \otimes \c{F}$ by the condition 
$a_l \ket{\bar i,0} = 0$ and we set recursively
$$
a^*_l\ket{\bar i,k_1,\dots ,k_l, \dots k_n}_F = 
\sqrt\kbar \sqrt{k_l+1}\,\ket{\bar i,k_1,\dots ,k_l+1, \dots k_n}_F.
$$
The coincidence limit is attained on elements of 
$L^2(V,d\mu) \otimes L^2(V,d\mu)$ of the form $\ket{\bar i,0}$.

The analogue of the integral $I$ defined in the Introduction is defined
then by the equation
$$
\bra{\bar j} G(q^\mu;q^{\nu\prime}) \ket{\bar i} = 
\bra{\bar j} I(\kbar\mu^2) \ket{\bar i}.
$$
Here $\mu$ is a parameter in the operator $\Delta$ with the dimension of
mass.  In general $I(\kbar\mu^2)$ is an operator acting on $\c{D}$. In
all the examples we shall consider however the space is homogeneous and
it reduces to a constant. We can write then
$$
\bra{\bar j} G(q^\mu;q^{\nu\prime}) \ket{\bar i} = 
I(\kbar\mu^2)\,\vev{\bar j\,|\,\bar i}.
$$
We represent this by the graph obtained by joining the ends of
(\ref{Feynman1}) and placing a $\bar j$ above and a $\bar i$ below the
circle which marks the join, as in the center of (\ref{Feynman2}) below. 

To calculate $\bra{\bar j} G(q^\mu;q^{\mu\prime}) \ket{\bar i}$
we must express $G$ in terms of the $a_l$ and their adjoints. For this 
we write
\be
\delta q^\mu = \sum_{l=1}^n (J^\mu_l a_l + J^{\mu *}_l a^*_l)     \label{na}
\ee
and from (\ref{diff-rel}) we conclude that
\be
\sum_{l=1}^n J^{[\mu}_l J^{\nu]*}_l = \frac 12 i q^{\mu\nu}.     \label{a-a}
\ee
The $J^\mu_l$ appear here as the components of a symplectomorphism.
They are fixed only to within a redefinition of the
$a_l$ and contain therefore $2n^2+n$ free parameters. This is the
number of elements of $GL(2n,\b{R})$ which leave invariant the
right-hand side of (\ref{a-a}).  If we interpret $\delta q^\mu$ as a
`string' joining two `points' $q^\mu$ and $q^{\mu\prime}$ then each
$a_j$ creates a longitudinal displacement. They would correspond to
the rigid longitudinal vibrational modes of the string. Since it
requires no energy to separate two points the string tension would be
zero. 

If the differential calculus $(\bar\Omega^*(\c{A}), \bar d)$ defined
in (\ref{d-bar})  has a frame $\bar \theta^\alpha = 
\bar \theta^\alpha_\lambda(\bar q^\mu)\, \bar d \bar q^\lambda$ then 
it would seem more appropriate to expand the variation in the form
\be
\bar \theta^\alpha_\lambda(\bar q^\mu) \, \delta q^\lambda = 
\sum_{l=1}^n (j^\alpha_l a_l + j^{\alpha *}_l a^*_l).           \label{cov}
\ee
We shall return to this Ansatz in Section~4. We are motivated here by
the desire to make $\delta q^\mu$ as similar as possible to the
element $dq^\mu$ of the differential calculus. This would suggest, in
particular, that the condition (\ref{b-d}) is fulfilled only if the
geometry is flat.

The `non-local' modification we shall find in the propagator is to
be associated not with the propagator but rather with the vertices at
its end points.  To see this we consider now the matrix elements
\bea
&&\hskip -1cm\bra{j,j^\prime} 
G(q^\mu;q^{\rho\prime}) 
\ket{i,i^\prime}\bra{l^\prime,l}
G(q^{\sigma\prime};q^\nu)
\ket{k^\prime,k} = \nonumber\\[4pt]
&&\hskip 1cm\bra{j}\otimes\bra{j^\prime}\otimes
\bra{l^\prime}\otimes\bra{l} 
G \otimes G \ket{i}\otimes\ket{i^\prime} \otimes
\ket{k^\prime}\otimes\ket{k} 
\eea
of the tensor product of two copies of the propagator, which we 
represent by the graph
\be
\begin{array}{ccccccc}
\hbox to0pt{\hss$j$\hss}
&&\hbox to0pt{\hss$j^\prime$\hss}
&&\hbox to0pt{\hss$l^\prime$\hss}
&&\hbox to0pt{\hss$l$\hss}\\[-2pt]
\circ&\hskip -.7em
\vrule height .6ex width 3em depth -.45ex\hskip-.7em&
\circ&\hskip 2em &\circ&\hskip -.7em
\vrule height .6ex width 3em depth -.45ex\hskip-.7em&\circ\\[-2pt]
\hbox to0pt{\hss$i$\hss}
&&\hbox to0pt{\hss$i^\prime$\hss}
&&\hbox to0pt{\hss$k^\prime$\hss}
&&\hbox to0pt{\hss$k$\hss}
\end{array}                                              \label{FFeynman1}
\ee
To form a vertex we must `join' the `point' 
$k^\prime$ 
to the `point' $i^\prime$. Following the prescription
(\ref{join}) this means that we replace the basis element
$$
\ket{i^\prime} \otimes \ket{k^\prime} \in
L^2(V,d\mu) \otimes L^2(V,d\mu)
$$
by the basis element
$$
\ket{\bar i^\prime} = \ket{\bar i^\prime, 0} \in
\c{D} \otimes \c{F}.
$$
We are prompted then to introduce the projection
$$
L^2(V,d\mu) \otimes L^2(V,d\mu) \otimes L^2(V,d\mu) \otimes L^2(V,d\mu)
\buildrel P \over \longrightarrow 
L^2(V,d\mu) \otimes \c{D} \otimes L^2(V,d\mu)
$$
defined by
$$
P = \sum_{r, \bar r^\prime, s} 
\ket{r, \bar r^\prime, s}\bra{r, \bar r^\prime, s} 
$$
and to define the propagator $G_2(q^\mu,q^{\rho\prime},q^\nu)$
in terms of the matrix elements
\bea
&&\hskip -1cm \bra{j, \bar j^\prime, l} G_2
\ket{i, \bar i^\prime, k} =\nonumber\\[8pt]
&&\sum_{r, \bar r^\prime, s} 
\bra{j, \bar j^\prime, l} G \otimes (1 \otimes 1)
\ket{r, \bar r^\prime, s}
\bra{r, \bar r^\prime, s} (1 \otimes 1) \otimes G
\ket{i, \bar i^\prime, k} =\nonumber\\[-2pt]
&&\hskip 1cm\sum_{r, \bar r^\prime, s}
\bra{j,\bar j^\prime} G \otimes 1 \ket{r, \bar r^\prime} \, \delta_{ls}
\delta_{ri} \,\bra{\bar r^\prime, s} 1 \otimes G
\ket{\bar i^\prime, k} =\nonumber\\[2pt]
&&\hskip 2cm\sum_{\bar r^\prime}
\bra{j,\bar j^\prime} G \otimes 1 \ket{i, \bar r^\prime}
\bra{\bar r^\prime, l} 1 \otimes G
\ket{\bar i^\prime, k}                                         \label{P2}
\eea
which we represent by the graph
\be
\begin{array}{ccccccc}
\hbox to0pt{\hss$j$\hss}
&&\hbox to0pt{\hss$\bar j^\prime$\hss}
&&\hbox to0pt{\hss$l$\hss}\\[0pt]
\circ&\hskip -.7em
\vrule height .6ex width 3em depth -.45ex\hskip-.7em&
\bigcirc&\hskip -.7em
\vrule height .6ex width 3em depth -.45ex\hskip-.7em&\circ
\\[0pt]
\hbox to0pt{\hss$i$\hss}
&&\hbox to0pt{\hss$\bar i^\prime$\hss}
&&\hbox to0pt{\hss$k$\hss}
\end{array}                                        \label{Feynman2}
\ee
We could have also included the dummy multiplication index and written
$$
\begin{array}{ccccccc}
\hbox to0pt{\hss$j$\hss}
&&\hbox to0pt{\hss$\bar j^\prime\;\bar r^\prime$\hss}
&&\hbox to0pt{\hss$l$\hss}\\[0pt]
\circ&\hskip -.7em
\vrule height .6ex width 3em depth -.45ex\hskip-.7em&
\bigcirc&\hskip -.7em
\vrule height .6ex width 3em depth -.45ex\hskip-.7em&\circ
\\[0pt]
\hbox to0pt{\hss$i$\hss}
&&\hbox to0pt{\hss$\bar r^\prime\;\bar i^\prime$\hss}
&&\hbox to0pt{\hss$k$\hss}
\end{array}  
$$
We have used the identifications
$$
G \otimes G = G \otimes (1 \otimes 1) \cdot 
(1 \otimes 1) \otimes G
$$
and the fact that $G \otimes G$ acts on
\bea
&&\hskip-1cm\Big(L^2(V,d\mu) \otimes L^2(V,d\mu)\Big) \otimes 
\Big((L^2(V,d\mu) \otimes L^2(V,d\mu)\Big) = \nonumber\\[2pt]
&&L^2(V,d\mu) \otimes \Big(L^2(V,d\mu) \otimes L^2(V,d\mu)\Big) 
\otimes L^2(V,d\mu) = \nonumber\\[2pt]
&&\hskip1cm L^2(V,d\mu) \otimes (\c{D} \otimes \c{F}) 
\otimes L^2(V,d\mu).\nonumber
\eea
Since $P$ projects $\c{D} \otimes \c{F}$ onto $\c{D}$ we see that
$$
G_2:\; L^2(V,d\mu) \otimes \c{D} \otimes L^2(V,d\mu)
\rightarrow  L^2(V,d\mu) \otimes \c{D} \otimes L^2(V,d\mu).
$$
In the commutative limit $\kbar\to 0$ one would find
$$
\bra{j, \bar j^\prime, l} G_2 \ket{i, \bar i^\prime, k} \to
G_2\, \delta_{ij}\delta_{i^\prime j^\prime}\delta_{kl}
$$
on the left-hand side of (\ref{P2}) and
$$
\bra{j,\bar j^\prime} G \otimes 1 \ket{i, \bar r^\prime} \to
G\, \delta_{ij}\delta_{r^\prime j^\prime}
$$
on the right-hand side. One would normally choose as basis the
eigenvectors of the position operator so that 
$q^\mu \ket{i} = q^\mu_i \ket{i}$ and one would normally drop the extra
index on $q^\mu$. The preceeding two limits would be written then
respectively 
$$
\bra{j, \bar j^\prime, l} G_2 \ket{i, \bar i^\prime, k} \to
G_2(q^\mu,q^{\rho\prime}, q^\nu)
$$
and
$$
\bra{j,\bar j^\prime} G \otimes 1 \ket{i, \bar r^\prime} \to
G(q^\mu,q^{\rho\prime}).
$$

The graph (\ref{Feynman2}) in turn can be cut into the two graphs 
\be
\begin{array}{ccccccc}
\hbox to0pt{\hss$j$\hss}
&&\hbox to0pt{\hss$\bar j^\prime$\hss}
&&\hbox to0pt{\hss$\bar r^\prime$\hss}
&&\hbox to0pt{\hss$l$\hss}\\[0pt]
\circ&\hskip -.7em
\vrule height .6ex width 3em depth -.45ex\hskip-.7em&
\subset&\hskip 2em &\supset&\hskip -.7em
\vrule height .6ex width 3em depth -.45ex\hskip-.7em&\circ
\\[0pt]
\hbox to0pt{\hss$i$\hss}
&&\hbox to0pt{\hss$\bar r^\prime$\hss}
&&\hbox to0pt{\hss$\bar i^\prime$\hss}
&&\hbox to0pt{\hss$k$\hss}
\end{array}                                            \label{ffeynman1}
\ee
which represent respectively the factors
$$
\bra{j,\bar j^\prime} G \otimes 1 \ket{i, \bar r^\prime}, \qquad
\bra{\bar r^\prime, l} 1 \otimes G
\ket{\bar i^\prime,k}.
$$
We are prompted by this to introduce also the graph
\be
\begin{array}{ccc}
\hbox to0pt{\hss$\bar j$\hss}
&&\hbox to0pt{\hss$\bar l$\hss}\\[0pt]
\supset&\hskip -.7em
\vrule height .6ex width 3em depth -.45ex\hskip-.7em&
\subset\nonumber\\[0pt]
\hbox to0pt{\hss$\bar i$\hss}
&&\hbox to0pt{\hss$\bar k$\hss}
\end{array}
\ee
to represent the matrix elements
$$
\bra{\bar j,\bar l} 1 \otimes G \otimes 1 \ket{\bar i, \bar k}.
$$
This is the propagator with `fuzzy' vertices. It is obtained by joining
$(i,j)$ to $(k,l)$ in the graph (\ref{Feynman2}) and cutting it as in 
(\ref{ffeynman1}). We designate it by $\bar G$:
$$
\bar G:\; \c{D} \otimes \c{D} \rightarrow \c{D} \otimes \c{D}
$$
If we replace the ends of (\ref{Feynman2}) by fuzzy vertices we obtain the
graph 
$$
\begin{array}{ccccccc}
\hbox to0pt{\hss$\bar j$\hss}
&&\hbox to0pt{\hss$\bar j^\prime$\hss}
&&\hbox to0pt{\hss$\bar l$\hss}\\[0pt]
\supset&\hskip -.7em
\vrule height .6ex width 3em depth -.45ex\hskip-.7em&
\bigcirc&\hskip -.7em
\vrule height .6ex width 3em depth -.45ex\hskip-.7em&\subset
\\[0pt]
\hbox to0pt{\hss$\bar i$\hss}
&&\hbox to0pt{\hss$\bar i^\prime$\hss}
&&\hbox to0pt{\hss$\bar k$\hss}.
\end{array} 
$$
This is a 2-line vertex. We designate it by $\bar G_2$:
$$
\bar G_2:\; \c{D} \otimes \c{D} \otimes \c{D} \rightarrow 
\c{D} \otimes \c{D} \otimes \c{D}.
$$

If we join the two ends we obtain a 2-line loop which we write also 
$\bar G_2$ but now
$$
\bar G_2:\; \c{D} \otimes \c{D} \rightarrow 
\c{D} \otimes \c{D}.
$$
Normally one vertex will be considered as fixed. If we trace over the
remaining one we shall use still the notation $\bar G_2$. We give a
simple example in the following section. 

The theory can be readily extended to incorporate a tree-level $n$-line
vertex. Consider as example a triple vertex. To pass from the equivalent
of (\ref{FFeynman1}) to (\ref{Feynman2}) we must replace (\ref{factor})
by the identification
$$
L^2(V,d\mu) \otimes L^2(V,d\mu) \otimes L^2(V,d\mu) \simeq
\c{D} \otimes \c{F} \otimes \c{F}
$$
which is obtained by introducing
$$
\bar q^\mu = \frac 13 (q^\mu \otimes 1 \otimes 1 + 
1 \otimes q^\mu \otimes 1 + 1 \otimes 1 \otimes q^\mu)
$$
as well as two Fock spaces to describe the variations. Three lines
are joined to a vertex by considering the tensor product of three
propagators:
$$
\bra{j,j^\prime} G \ket{i,i^\prime}
\bra{l,l^\prime} G \ket{k,k^\prime}
\bra{n,n^\prime} G \ket{m,m^\prime}
$$
and projecting the element
\bea
&&\hskip -1cm\ket{i}\otimes\ket{i^\prime}\otimes
\ket{k}\otimes\ket{k^\prime}\otimes
\ket{m}\otimes\ket{m^\prime} \in\nonumber\\[2pt] 
&&\hskip 1cmL^2(V,d\mu) \otimes L^2(V,d\mu) \otimes L^2(V,d\mu)
\otimes L^2(V,d\mu) \otimes L^2(V,d\mu) \otimes L^2(V,d\mu)  \nonumber
\eea
onto an element 
$$
\ket{i}\otimes
\ket{k}\otimes
\ket{m}\otimes\ket{\bar i^\prime} \in 
L^2(V,d\mu) \otimes L^2(V,d\mu) \otimes L^2(V,d\mu) \otimes \c{D}
$$
The way this projection is defined will depend on which lines are to be
considereed as incoming and which are outgoing. The above construction
of joining and cutting would lead to the vextex defined by the matrix
elements
$$
\bra{\bar j,\bar l,\bar n} G_3(q^\mu_1;q^\mu_2;q^\mu_3) 
\ket{\bar i,\bar j,\bar m}.
$$

\sect{The noncommutative flat plane}

The noncommutative flat plane is the algebra $\c{A}_\kbar$ generated by
two hermitian elements $q^1 = x$ and $q^2 = y$ which satisfy the
commutation relation $[x,y] = i \kbar$ and which has over it the
differential calculus $\Omega^*(\c{A}_\kbar)$ given by 
$[q^\mu, dq^\nu] = 0$. If we introduce the two derivations
$$
e_1 = - {1\over i\kbar} \ad y, \qquad
e_2 =  {1\over i\kbar} \ad x
$$
dual to $dq^\mu$ then an appropriate generalization~\cite{MadMou98} of
the Laplace operator $\Delta$ with mass $\mu$ is given by
$$
\Delta = \Delta_\kbar +\mu^2, \qquad \Delta_\kbar = - (e^2_1 + e^2_2).
$$
For each couple $(k_1,k_2) \in \b{R}^2$ we introduce the unitary elements 
$u(k_1), \; v(k_2) \in \c{A}_\kbar$ defined by
$$
u(k_1) = e^{ik_1 x}, \qquad v(k_2) = e^{ik_2 y}.
$$
They satisfy the commutation relations 
$$
u(k_1)v(k_2) = q^{k_1k_2\kbar} v(k_2)u(k_1), \qquad q = e^{-i}.
$$
A basis for the Hilbert space $\c{H}$ is given by the eigenvectors
$$
\phi_k = u(k_1) v(k_2), \qquad k = (k_1, k_2)
$$
of $\Delta$.  The corresponding eigenvalues are 
$$
\lambda_k = k^2 + \mu^2, \qquad k^2 = k^2_1 + k^2_2.
$$
The propagator can be written then
$$
G(x,y;x^\prime,y^\prime) =
\frac{1}{(2\pi)^2} \int (k^2 + \mu^2)^{-1}
\phi^\prime_k \otimes 
\phi^*_k \, dk, \qquad dk = dk_1 dk_2.
$$
We must introduce a partial trace on $\c{A}_\kbar$. This can be done only
through a representation.  The only properties which we shall need
however are the identities
$$
\tr(u^*(k^\prime_1)u(k_1)) = 
2\pi \delta(k^\prime_1 - k_1), \qquad
\tr(v^*(k^\prime_2)v(k_2)) = 
2\pi \delta(k^\prime_2 - k_2).
$$
That is:
$$
\tr(\phi^*_{k^\prime} \phi^{\phantom{*}}_k) = 
(2\pi)^2\delta^{(2)}(k^\prime - k).
$$

The commutation relations (\ref{diff-rel}) become in this case
\be
[\bar x, \bar y] = \frac 12 i \kbar,   \qquad
[\delta x, \delta y] = \frac 12 i \kbar.                    \label{half}
\ee
As in (\ref{na}) we write
\be
\delta x = J^1 a + J^{1*} a^*, \qquad
\delta y = J^2 a + J^{2*} a^*.                               \label{a}
\ee
With (\ref{a-a*}) satisfied we have $J^{[1} J^{2]*} = \frac 12 i q^{12}$.
By a redefinition of $a$ we can choose
$$
J^1 = \frac 12, \qquad J^2 = \frac 1{2i}, \qquad a = \delta x + i \delta y
$$
The freedom here is $SL(2, \b{R})$, the symplectomorphism group in 
dimension 2. By a renormalization of $\kbar$ we can also choose 
$q^{12} = 1$.

We index the basis of $L^2(V,d\mu) = L^2(\b{R}^2,dp)$ by 
$p = (p_1, p_2)$ and introduce the basis 
$\ket{\bar p,k} = \ket{\bar p}_D \otimes \ket{k}_F$ according to the
prescription (\ref{factor}) of the previous section. We shall also
re-express the tensor product according to (\ref{bar-delta}) and drop
the tensor-product symbol. We have then
$$
u^{\prime *}(k_1)\ket{\bar p^\prime} = e^{-ik_1 x^\prime}\ket{\bar p^\prime} = 
e^{-ik_1(\bar x + \delta x)}\ket{\bar p^\prime}.
$$
Since $\bar x$ and $\delta x$ commute we can write this as
$$
u^{\prime *}(k_1)\ket{\bar p^\prime} = 
e^{-ik_1 \bar x} e^{-ik_1(a + a^*)/2} \ket{\bar p^\prime}.
$$
Using the Baker-Campbell-Hausdorff (BaCH) formula
$$
e^{\alpha a + \beta a^*} = 
e^{\beta a^*} e^{\alpha a} e^{\alpha\beta \kbar/2} = 
e^{\alpha a} e^{\beta a^*} e^{-\alpha\beta\kbar/2}
$$
we find that
$$
u^{\prime *}(k_1)\ket{\bar p^\prime} = 
e^{-ik_1 \bar x} e^{-k_1^2 \kbar/8} e^{-ik_1a^*/2}\ket{\bar p^\prime}
$$
and therefore
\bea
\phi^{\prime *}_k\ket{\bar p^\prime} \squeeze &&= 
e^{-ik_2y^\prime} e^{-ik_1 \bar x} e^{-k_1^2\kbar/8}
e^{-ik_1 a^*/2} \ket{\bar p^\prime} \nonumber\\[6pt]
&&= e^{-ik_2\bar y} e^{-ik_1\bar x} e^{-\kbar k^2/8}
e^{k_2 a^*/2}e^{-k_2 a/2} e^{-ik_1 a^*/2}\ket{\bar p^\prime}  \nonumber\\[6pt]
&&= e^{-ik_2\bar y} e^{-ik_1\bar x} e^{-\kbar k^2/8}
e^{ik_1k_2\kbar/4} e^{(k_2 - ik_1) a^*/2}\ket{\bar p^\prime}.  \nonumber
\eea
Similarly we find
$$
\phi^*_k\ket{\bar p} = 
e^{-ik_2\bar y} e^{-ik_1\bar x} e^{-\kbar k^2/8}
e^{ik_1k_2\kbar/4}e^{-(k_2 - ik_1) a^*/2}\ket{\bar p}.
$$
{From} these last two equations we deduce that
\be
\bra{\bar p^\prime}\phi^{\phantom{*}}_k \otimes\phi^*_k\ket{\bar p}
= e^{-\kbar k^2/2} \vev{\bar p^\prime\,|\, \bar p}. \label{flat-Green}
\ee
The product here is the tensor product (\ref{bar-delta}). Since the
$\bar \c{A}_\kbar$ factor reduces in fact to the identity, the product 
depends only on the second factor $\delta\c{A}_\kbar$. We have dropped
the prime on $\phi_k$ since the information is contained in the position
in the tensor product. 

The Fourier transform is the map 
\be
\tilde \phi(k) = {1 \over (2\pi)^2} \tr (\phi^*_k \,\phi)         \label{f}
\ee
from $\c{H}$ to the momentum space $L^2(\b{R}^2,dk)$ and the map 
\be
\phi = \int \phi_k \,\tilde\phi (k) dk =
\int e^{ik_2y} e^{ik_1x} e^{-ik_1k_2\kbar}\,\tilde\phi (k) dk     \label{i-f}
\ee
from $L^2(\b{R}^2,dk)$ to $\c{H}$. The Plancherel theorem is
the completeness relation for the set of $\phi_k$. We have the 
unitary map 
$$
\tilde \phi(l) = 
{1 \over (2\pi)^2} \tr \Big(\phi^*_l 
\int \tilde\phi(k) \,\phi_k \,dk\Big)
$$
from $L^2(\b{R}^2,dk)$ onto itself and the unitary map
$$
\phi \otimes 1 = {1 \over (2\pi)^2} \int 
\tr_2(\phi^{\phantom{*}}_k \otimes 
\phi^*_k \cdot 1 \otimes \phi)
$$
of $1 \otimes \c{H}$ onto $\c{H} \otimes 1$. Introduce
$$
\tilde D_1 = \tilde \partial_1, \qquad
\tilde D_2 = \tilde \partial_2 - i k_1\kbar, \qquad 
[\tilde D_1, \tilde D_2] = - i \kbar.
$$
The multiplication by $x$ and $y$ are transformed respectively into
the operators $i\tilde D_1$ and $i\tilde D_2$, which are self-adjoint
on $L^2(\b{R}^2,dk)$. The Fourier transform respects the commutation
relations. The multiplication by $x \pm i y$ are transformed
respectively into $\tilde b$ and $\tilde b^*$ where
$$
\tilde b = i \tilde \partial_1 - \tilde \partial_2 + i \kbar k_1.
$$
The asymmetry in the Fourier transform of the multiplication
operators is due to our convention in the choice of basis $\phi_k$. In 
the analogous calculations in the quantum Hall effect one would speak
of a choice of gauge. If one introduces the `gauge symmetric' operators
\bea
&&\tilde b^\prime = e^{\kbar k^2/2} \,\tilde b \,e^{-\kbar k^2/2} = 
i \tilde \partial_1 - \tilde \partial_2 + \frac 12 \kbar (ik_1 + k_2)
\nonumber\\[4pt]&&
\tilde b^{\prime *} = e^{-\kbar k^2/2} \,\tilde b^* \,e^{\kbar k^2/2} = 
i \tilde \partial_1 + \tilde \partial_2 + \frac 12 \kbar (ik_1 + k_2)
\nonumber
\eea
then $\tilde b^{\prime *}$ is the adjoint of $\tilde b^\prime$ on
$L^2(\b{R}^2, e^{-\kbar k^2}dk)$. This symmetric form emphasizes 
the role of the commutation relations in position space as a cut-off
in momentum space.

The Fourier transform defines the map 
$$
\tilde \phi(k, k^\prime) = 
{1 \over (2\pi)^4} \tr (\phi^*_k \otimes \phi^*_{k^\prime} \,
\phi \otimes \phi) = 
{1 \over (2\pi)^4} \tr (\phi^*_k \,\phi)\, 
\tr (\phi^*_{k^\prime} \,\phi)
$$
from $\c{H} \otimes \c{H}$ to $L^2(\b{R}^2,dk) \otimes L^2(\b{R}^2,dk)$ 
and the map 
$$
\phi\otimes\phi = \int \phi_k \otimes \phi_{k^\prime}\,
\tilde\phi (k,k^\prime) dk dk^\prime
$$
from $L^2(\b{R}^2,dk) \otimes L^2(\b{R}^2,dk)$ to $\c{H}\otimes\c{H}$. 
If we write $\phi \otimes \phi = \bar \phi \otimes \delta \phi$
as in (\ref{bar-delta}) then (\ref{flat-Green}) states that the
Fourier transform of the diagonal factor of 
$\phi_k \otimes \phi_{k^\prime}$ is a constant function and that the
projection onto the ground-state in $\c{F}$ produces an exponential 
damping in momentum space.

We are now in a position to calculate the coincidence limit of the 
propagator. We have
\bea
\hskip -1cm\bra{\bar p^\prime}G(x,y;x^\prime,y^\prime)\ket{\bar p} 
\squeeze &&= {1\over (2\pi^2)} \int (k^2 + \mu^2)^{-1}
\bra{\bar p^\prime}\phi^{\phantom{*}}_k 
\otimes \phi^*_k \ket{\bar p} dk \nonumber\\[6pt]
&&= {1\over (2\pi^2)} \int {e^{-\kbar k^2/2} \over k^2 + \mu^2}\,
\vev{\bar p^\prime\,|\, \bar p} dk.\nonumber  
\eea
The Feynman rules here are the same as the commutative ones except for 
an extra factor $e^{-\kbar k^2/4}$ at each end of a propagator of
momentum $k$ to account for the projection onto the ground state in
$\c{F}$. We find then
$$
\bra{\bar p}G(x,y;x^\prime,y^\prime)\ket{\bar p}
\nonumber\\[4pt] = I(\kbar \mu^2)\, \vev{\bar p \,|\, \bar p}
$$
where $I(\kbar \mu^2)$ is given by the 
integral~\cite{PaiUhl50, DopFreRob95, ChaDemPre98}
\be
I(\kbar \mu^2) =
\frac{1}{(2\pi)^2} \int {e^{-\kbar k^2/2} \over k^2 + \mu^2} dk.  \label{I}
\ee
With a change of variable it can be written as
$$
I(\kbar\mu^2) = \frac 1{4\pi}
\int_0^\infty {e^{-x} \over x + \kbar \mu^2/2} dx
= - \frac 1{4\pi} e^{\kbar\mu^2/2} \mbox{Ei}(-\kbar\mu^2/2),
$$
where $\mbox{Ei}(x)$ is the exponential-integral function.
When $\kbar\mu^2 \to 0$ one finds
$$
I(\kbar\mu^2) =  \frac 1{4\pi} 
\Big( - \log (\kbar\mu^2) + \log 2 - \gamma - 
\frac 12 \kbar\mu^2 \log (\kbar\mu^2) +
o( \kbar \mu^2)\Big)
$$
and when $\kbar\mu^2 \to \infty$,
$$
I(\kbar\mu^2) = \frac 1{2\pi\kbar\mu^2} 
+ o((\kbar\mu^2)^{-2}).
$$

As a further illustration of the modified Feynman rules, we calculate
the 2--point function obtained by integrating over the internal vertex in 
$$
G_2(q^{\mu};\bar{q}^{\rho\prime};q^{\nu}) 
\in \c{A}_\kbar \otimes \bar \c{A}_\kbar \otimes \c{A}_\kbar,
$$
represented by (\ref{Feynman2}).  In terms of the eigenfunctions of the
Laplacian, the definition (\ref{P2}) of
$G_2(q^{\mu};\bar{q}^{\rho\prime};q^{\nu})$ can be written as
$$
G_2(q^{\mu};\bar{q}^{\rho\prime};q^{\nu}) = 
\frac 1{(2\pi)^4} \int dk dl \lambda_k^{-1} \lambda_l^{-1}
\phi_l \otimes {}_F\bra{0} \phi_l^* \otimes \phi_k \ket{0}_F 
\otimes \phi_k^*.
$$
The ${}_F\bra{0} \phi_l^* \otimes \phi_k \ket{0}_F$ is the
projection onto the ground state in $\c{F}$.  Integration over 
$\bar q^{\rho\prime}$ corresponds to taking the trace over 
$\bar \c{A}_\kbar$. Similarly to (\ref{flat-Green}), it is
straightforward to calculate
\bea
\tr_{\bar \c{A}_\kbar} 
        ({}_F\bra{0} \phi_l^* \otimes \phi_k\ket{0}_F) 
  &=&  \tr(e^{-il_2\bar y}e^{i(k_1-l_1)\bar x}e^{ik_2 \bar y})\ 
       {}_F\bra{0} e^{ik_2 \delta y} 
       e^{2ik_1 \delta x} e^{ik_2 \delta y} \ket{0}_F \nonumber \\
  &=&  (2\pi)^2 \delta^{(2)}(k-l) {}_F\bra{0} e^{ik_2 \delta y} 
       e^{2ik_1 \delta x} e^{ik_2 \delta y} \ket{0}_F \nonumber \\
  &=&  (2\pi)^2 \delta^{(2)}(k-l) e^{-\kbar k^2/2}.  \label{trace_calc}
\eea
Therefore 
\be
\tr_{\bar \c{A}_\kbar} (G_2(q^{\mu};\bar{q}^{\rho\prime};q^{\nu})) = 
   (2\pi)^{-2} \int dk \frac{e^{-\kbar k^2/2}} {(k^2+\mu^2)^2} 
   \phi_k\otimes \phi_k^*.
\ee
Again, we see that the Feynman rules are the same as in the commutative
case, except for an extra factor $e^{-\kbar k^2/4}$ at the 
end of a propagator of momentum $k$. Similarly for higher order vertices,
the projection onto the ground state of the relative coordinates
will lead to an exponential damping factor with length scale set by $\kbar$,
since the plane-wave factors as in (\ref{trace_calc}) act as unitary 
operators which shift $\ket{0}_F$, reducing the overlap with 
${}_F\bra{0}$.

As a second example consider a 2-line loop with no momentum flowing
through it:
\bea
&&\bra{\bar p\bar p} \bar G_2(q^\mu;q^{\nu\prime})
\ket{\bar p \bar p} =\nonumber\\[4pt]
&&\hskip 2cm\int \bra{\bar p\bar p} \bar G(q^\mu;q^{\nu\prime})
\ket{\bar p^\prime \bar p^\prime}
\bra{\bar p^\prime,\bar p^\prime} 
\bar G(q^\mu;q^{\nu\prime})
\ket{\bar p,\bar p} d\bar p^\prime. \nonumber
\eea
If we set as before 
$$
\mu^2 \bra{\bar p,\bar p}
\bar G_2(q^\mu;q^{\nu\prime})\ket{\bar p, \bar p}
 = I_2(\kbar\mu^2) \vev{\bar p,\bar p\,|\,\bar p,\bar p}
$$
then we find that
$$
I_2(\kbar\mu^2) = \frac {\mu^2}{(2\pi)^4} 
\int {e^{-\kbar k^2} \over (k^2 + \mu^2)^2} dk
= \frac 1{32\pi^3} + \frac {\kbar\mu^2}{8\pi^3}
e^{\kbar\mu^2} \mbox{Ei}(-\kbar\mu^2).
$$
When $\kbar\mu^2 \to 0$
$$
I_2(\kbar\mu^2) =  \frac 1{16\pi^3}\Big(1 + 
\kbar \mu^2 \log (\kbar \mu^2) + \cdots\Big)
$$
and when $\kbar\mu^2 \to \infty$
$$
I_2(\kbar\mu^2) = \frac 1{32\pi^3 \kbar\mu^2} 
\Big(1 - \frac 2{\kbar\mu} + \cdots\Big).
$$
It is remarkable that this vanishes to the same order in
$(\kbar\mu^2)^{-1}$ as $I(\kbar\mu^2)$ when $\kbar\mu^2 \to \infty$. 

We have represented only the difference $\delta q^\mu$ in terms
of annihilation and creation operators. It is possible to represent also
$\bar q^\mu$.  We shall argue below that this is necessary on a curved 
noncommutative geometry. For this we introduce as well as $a$ defined in
(\ref{a}) the operator $b = \bar x + i \bar y$.  Then it is easy to
see that the commutation relations (\ref{a-a*}) hold also for $b$ and
that $a$ and $b$ commute with each other and their adjoints. We define
as usual the commuting number operators
$$
N_{a} = \frac{1}{\kbar}a^* a , \qquad N_{b} = \frac{1}{\kbar}b^* b
$$
and let $\ket{n_a , n_b}$ be their common eigenvectors.
The equations which defined $a$ and $b$ can be inverted to yield
$$
\begin{array}{ll}
\displaystyle{x \otimes 1 = \frac{1}{2}(- a - a^* + b + b^* )},
&\displaystyle{y \otimes 1 = \frac{1}{2i}(- a + a^* + b - b^* )}, \\[12pt]
\displaystyle{1 \otimes x =  \frac{1}{2}(a + a^* + b + b^* )},
&\displaystyle{1 \otimes y = \frac{1}{2i }(a - a^* + b - b^* )}
\end{array}
$$
and therefore we find
$$
\begin{array}{ll}
u (k_{1}) = e^{i k_{1} (-a - a^* + b + b^*)/2 },
&u^{\prime*} (k_{1}) = e^{-i k_{1} (a + a^* + b + b^*)/2 }, \\[6pt]
v (k_{2}) = e^{k_{2} (-a + a^* + b - b^*)/2 },
&v^{\prime*} (k_{2}) = e^{- k_{2} (+a - a^* + b - b^*)/2 }.
\end{array}
$$
Now it is straightforward to show that Equation~(\ref{flat-Green}) can be
written more generally as
$$
\phi^{\phantom{*}}_k \otimes \phi^*_k
  = e^{-(k_2 + ik_1 )\,a + (k_2 - ik_1 ) \,a^* }.
$$
The $b$-terms cancel. Using the BaCH formula we
find that
$$
e^{-(k_2 + ik_1 )a + (k_2 - ik_1 ) a^* } 
= e^{-\kbar k^2/2} e^{(k_2 - ik_1 )a^* }e^{-(k_2 + ik_1 )a }.
$$
Thus we obtain
\bea
&&\bra{m_a , m_b}\phi^{\phantom{*}}_k \otimes \phi^*_k
\ket{ n_a ,n_b} \nonumber\\[6pt]
&&\hskip 2cm = e^{-\kbar k^2/2}
\bra{m_a}e^{(k_2 - ik_1 )a^* }e^{-(k_2 + ik_1 )a } 
\ket{ n_a}\delta_{m_b n_b }. \nonumber
\eea
>From the expansion
$$
e^{\alpha a } \ket{n_a} = \ket{n_a} + 
\alpha \sqrt{\kbar}\sqrt{n_a} \ket{n_a - 1} + \cdots + 
\frac{(\alpha \sqrt{\kbar})^{n_a} }{ n_a ! }\sqrt{n_a ! } \ket{0}.
$$                                                           
it follows then that
$$
\bra{m_a}e^{(k_2 - ik_1 )a^*}e^{-(k_2 + ik_1 )a } \ket{n_a}
$$
can be calculated for any two given states.  We are especially
interested in the case when $m_a = n_a $. In this case
\bea
&&\bra{n_a , m_b}\phi^{\phantom{*}}_k \otimes \phi^*_k 
\ket{n_a,n_b} \nonumber\\[6pt]
&&\hskip 2cm = e^{-\kbar k^2/2} \Big( 1 - \kbar k^2 n_a + \cdots + 
\frac{(-\kbar)^{n_a}(k^2)^{n_a} }{ (n_a !)^2 } n_a !\Big) \delta_{m_b n_b}.
\nonumber
\eea
The propagator is given therefore by
\bea
&& \bra{n_a , m_b} G(x, y ; x', y' ) \ket{n_a , n_b}  \nonumber\\[6pt]
&& \hskip 1cm = \frac{1}{(2\pi)^2 } 
\int {e^{-\kbar k^2/2} \over k^2 + \mu^2} 
\Big(1 - \kbar k^2 n_a + \cdots  + 
\frac{(-\kbar)^{n_a}(k^2)^{n_a} }{ (n_a !)^2 } n_a !\Big) 
dk\, \delta_{m_b n_b }.\nonumber
\eea
This equation generalizes Equation~(\ref{I}) for $I(\kbar\mu^2)$
and reduces to it when $n_a = 0$.

A more elegant formulation can be given with a more explicit
use~\cite{GroPre93, KemManMan95, KemMan97, ChaDemPre98} of the coherent
state formalism. Define
$$
a = \frac{1}{\sqrt{2}} (x + iy)\otimes 1, \qquad
b = \frac{1}{\sqrt{2}} \, \, 1 \otimes (x +iy).
$$
Then again it is easy to see that the commutation relations (\ref{a-a*})
hold for $a$ and $b$ and that $a$ and $b$ commute with each other and
their adjoints. Introduce, for $\tilde x, \tilde y \in \b{R}$
$$
z = \frac{1}{\sqrt{2} \kbar}(\tilde{x} + i\tilde{y}), \qquad
T(z) = e^{z a^* - \bar{z}a } 
$$
and similarly for $b$. Then the coherent states are given by 
$\ket{z} = T(z)\ket{0}$.  It is straightforward to see that
$a\ket{z} = \kbar  z \ket{z}$ and that $(x,y)$ are related to 
$(\tilde x, \tilde y)$ by
$\bra{z} x \ket{z} = \tilde{x}$, $\bra{z} y \ket{z} = \tilde{y}$. 
We argued above that we can express the variations $\delta x$ and
$\delta y$ using the tensor product of two copies of the algebra. Since 
\bea
&&\bra{z} \phi_k \ket{z} = e^{- i\kbar k_1 k_2/2 }
e^{-\kbar k^2/4}
e^{-i(k_1 \tilde x + k_2 \tilde y )},\nonumber\\[6pt]
&&\bra{z^\prime} \phi^{\prime*}_k \ket{z^\prime} = 
e^{i \kbar k_1 k_2/2} e^{-\kbar k^2/4}
e^{i(k_1 \tilde x^\prime + k_2 \tilde y^\prime )},\nonumber
\eea
we obtain
$$
\bra{z, z^\prime}\phi^{\phantom{*}}_k \otimes \phi^*_k 
\ket{z, z^\prime} = e^{-\kbar k^2/2} 
e^{i k_1 (\tilde x^\prime - \tilde x) + 
i k_2 (\tilde y^\prime -\tilde y)},
$$
and therefore
$$
\bra{z,z^\prime} G(x,y;x^\prime,y^\prime) \ket{z,z^\prime} =
\frac 1{4\pi^2} \int {e^{-\kbar k^2/2} \over k^2 + \mu}
e^{i k_1 (\tilde x^\prime - \tilde x) + 
i k_2 (\tilde y^\prime -\tilde y)}dk.
$$
When $(\tilde x^\prime, \tilde y^\prime) \rightarrow (\tilde x, \tilde y)$ 
it follows that
$$
\bra{z,z^\prime} G(x,y;x^\prime,y^\prime) 
\ket{z,z^\prime} \to I(\kbar\mu^2).
$$

The results we have obtained using only the abstract algebraic
structure of the noncommutative flat plane can be of course found also
using a specific representation. One such is the standard irreducible
representation of $\c{A}_\kbar$ as an $\mbox{I}_\infty$ factor on
$L^2(\b{R},d\alpha)$ given on $f(\alpha) \in L^2(\b{R},d\alpha)$ by
$$
u(k_1)f(\alpha) = e^{ik_1\alpha} f(\alpha), \qquad
v(k_2)f(\alpha) = f(\alpha + k_2 \kbar).
$$
A convenient basis for $L^2(\b{R},d\alpha)$ is $\ket{p_1} = e^{ip_1 \alpha}$
with $p_1\in \b{R}$. We have then
$$
u(k_1)\ket{p_1} = \ket{p_1 + k_1}, \qquad
v(k_2)\ket{p_1} = e^{ip_1k_2\kbar}\ket{p_1}.
$$
The parameter $p_1$ can be thought of as the momentum conjugate to $x$ but
this fact plays no role here. The eigenvectors 
$\phi_k = u(k_1) v(k_2)$ have matrix elements defined by
$$
\phi_k\ket{p_1} = e^{ip_1k_2\kbar}\ket{p_1+k_1}.
$$
This representation has a bad `classical' limit. The generator $x$
can be identified then with the parameter $\alpha$ but the generator $y$
tends to zero as $\kbar \to 0$. To obtain a sensible classical limit one
needs two copies of $L^2(\b{R},d\alpha)$. To see this we
define $u(k_1)$ and $v(k_2)$ on 
$L^2(\b{R}^2,d\alpha\,d\beta)$ as the operators
\bea
&&(u(k_1)f)(\alpha,\beta) = e^{i(a k_1 \alpha + b k_1 \beta)}
f(\alpha + c k_1 \kbar, \beta + d k_1 \kbar), \nonumber\\[4pt]
&&(v(k_2)f)(\alpha,\beta) = e^{i(a^\prime k_2 \alpha + b^\prime k_2 \beta)} 
f(\alpha + c^\prime k_2 \kbar, \beta + d^\prime k_2 \kbar)      \label{rep}.
\eea
If we choose $a+b=1$, $a^\prime + b^\prime=1$, we obtain a
representation with a non-degenerate limit with
$$
u(k_1)v(k_2) = q^{(d - c^\prime)k_1 k_2\kbar} v(k_2)u(k_1).
$$  
We can conclude then that if $d - c^\prime = 1$ one obtains a
representation of the algebra. We conclude also that if $c^\prime = d$
then $u$ and $v$ commute. The representation is therefore not
irreducible since the commutant is non-trivial.  We shall choose
$$
a = 1, \qquad a^\prime = 0,\qquad b = 0, \qquad b^\prime = 1.
$$

The propagator can be calculated directly in any one of the
representations (\ref{rep}). One obtains
$$
(\phi_k f)(\alpha,\beta) =
e^{id k_1k_2\kbar} e^{i (k_1 \alpha + k_2 \beta)}
f (\alpha + ck_1\kbar + c^\prime k_2\kbar,\,
\beta + dk_1\kbar + d^\prime k_2\kbar)
$$
and therefore
\bea
&&\hskip -0.5cm(\phi^{\phantom{*}}_k \otimes \phi^*_k \, 
f \, f^\prime) (\alpha,\beta;\alpha^\prime,\beta^\prime) =
e^{-ik_1k_2\kbar} e^{i (k_1 (\alpha - \alpha^\prime) + 
k_2 (\beta - \beta^\prime))} \times \nonumber\\[4pt]&&\hskip 0.5cm 
f (\alpha + ck_1\kbar + c^\prime k_2\kbar,\,
\beta + dk_1\kbar + d^\prime k_1\kbar) \,
f^\prime (\alpha^\prime - ck_1\kbar - c^\prime k_2\kbar,\,
\beta^\prime - dk_1\kbar - d^\prime k_1\kbar).\nonumber
\eea
Consider in particular the `plane-wave' basis 
$\ket{p} = e^{ip_1\alpha + ip_2\beta}$ of 
$L^2(\b{R}^2,d\alpha \,d\beta)$. Then we find
$$
\phi_k \ket{p} = 
e^{idk_1k_2\kbar}e^{ip_1(ck_1 + c^\prime k_2)\kbar}\,
e^{ip_2(dk_1 + d^\prime k_1)\kbar}\ket{p + k}
$$
and therefore
\bea
&& \hskip -1cm\phi^{\phantom{*}}_k\otimes \phi^*_k 
\ket{p;p^\prime} = \nonumber\\[4pt]
&& \hskip 1cm e^{-ik_1k_2\kbar}
e^{i(p_1 - p_1^\prime)(ck_1 + c^\prime k_2)\kbar}
e^{i(p_2 - p_2^\prime)(dk_1 + d^\prime k_1)\kbar}
\ket{p + k;p^\prime - k}.\nonumber
\eea
We are interested in the limit $p^\prime \to p$:
$$
\phi^{\phantom{*}}_k\otimes \phi^*_k 
\ket{p;p} = e^{-ik_1k_2\kbar}\ket{p + k;p - k}.
$$

The decomposition (\ref{bar-delta}) of the tensor product 
$\c{A}_\kbar \otimes \c{A}_\kbar$ is equivalent to a reparametrization
of the tensor product
$L^2(\b{R}^2,d\alpha\,d\beta) \otimes
L^2(\b{R}^2,d\alpha^\prime\,d\beta^\prime)$ induced by the linear
transformation
$$
(\alpha,\,\beta,\,\alpha^\prime,\,\beta^\prime) \to
(\frac 12 (\alpha^\prime + \alpha),\,\frac 12 (\beta^\prime + \beta),\, 
\alpha^\prime - \alpha,\,\beta^\prime - \beta)
$$
of the parameter space. The first two of the new coordinates yield the 
representation space $\c{D}$ of $\bar x$ and $\bar y$ and the second
two the representation space $\c{F}$ of $\delta x$ and $\delta y$.

The basis given above for the representation space is singular and it
is appropriate to change it, at least for the factor $\c{F}$. This 
is equivalent to the introduction of a form factor 
$F(\alpha^\prime - \alpha, \beta^\prime - \beta)$. For each choice of
$F$ we introduce $I_F(\kbar\mu^2)$ defined by the equation
$$
\bra{p;p} G(x,y;x^\prime, y^\prime) \,F \ket{p;p}
 = I_F(\kbar\mu^2) \bra{p;p} F \ket{p;p}.
$$
The `coherent-state' basis has a fundamental cell of
minimal area and the distance between two closest `points' is
minimal. So normally  one might expect that every choice of $F$ would
yield a value of $I_F(\kbar\mu^2)$ strictly less than 
$I(\kbar\mu^2)$. However this is not the case. For example a sequence
of $F$ which tends to the product of two $\delta$-functions,
$$
F \to \delta(\alpha^\prime - \alpha)\, \delta(\beta^\prime - \beta),
$$
will yield a value of $I_\delta (\kbar\mu^2)$ which is smaller than
$I(\kbar\mu^2)$ for sufficiently small values of $\kbar\mu^2$.
We obtain in fact
$$
I_\delta(\kbar\mu^2) = 
\frac 1{4\pi^2} \int {e^{ik_1k_2\kbar} \over k^2 + \mu^2} dk.
$$
A change of variables yields the expression
\be
I_\delta(\kbar \mu^2) =
\frac{1}{4\pi} \int_0^\infty {J_0(x) \over x + \kbar\mu^2/2}dx =
\frac 18 ({\bf H}_0(\kbar\mu^2/2) - Y_0(\kbar\mu^2/2)).             \label{I2}
\ee
Here ${\bf H}_0$ is a Struve function and $Y_0$ is a Neumann
function. When $\kbar\mu^2 \to 0$ one finds
$$
I_\delta(\kbar\mu^2) =  \frac 1{4\pi} 
\Big( - \log (\kbar\mu^2) + 2\log 2 - \gamma + 
o(\kbar\mu^2)\Big)
$$
and when $\kbar\mu^2 \to \infty$,
$$
I_\delta(\kbar\mu^2) = \frac 1{2\pi\kbar\mu^2} +
o((\kbar\mu^2)^{-2}.
$$
Comparing the two asymptotic expansions we find
\bea
&&I(\kbar\mu^2) - I_\delta(\kbar\mu^2) = - \frac 1{8\pi} \Big(\log 2 
+ \frac 12 \kbar\mu^2 \log (\kbar\mu^2) + o(\kbar \mu^2)\Big),\nonumber \\ 
&&I(\kbar\mu^2) - I_\delta(\kbar\mu^2) = 
- {1 \over 2\pi(\kbar \mu^2)^2} + o((\kbar \mu^2)^{-3}).\nonumber
\eea
The two functions agree to the dominant term in $\kbar\mu^2$ for large
and small values but at least to the sub-dominant terms it is rather
$I(\kbar\mu^2)$ which is the smaller.

The modification of the propagator which we have found is due to the
noncommutativity of the algebra. However we saw that we could define
the variation of an element of a noncommutative algebra using the
tensor product of two copies of it.  The effect then was formally
encoded in the difference between a product and a tensor product; the
generator $x$ does not commute with $y$ but it does commute with
$y^\prime$. In a subsequent article we shall discuss also a braided
tensor product, which has all of the properties of an ordinary
product. Although it is somewhat formal, one could consider an analog
in the present situation by setting also $[x,y^\prime] = i\kbar$. In
this case the properties of the variation of an element would not be
correctly encoded in the tensor product. One would find that the
commutation relations (\ref{half}) were in fact replaced by
$$
[\bar x, \bar y] = i \kbar,   \qquad
[\delta x, \delta y] = 0.
$$
The noncommutative propagator is seen to be exactly the classical
propagator.  The propagator depends, we have seen, only on the
variations $\delta x$ and $\delta y$.

The self-energy of a scalar particle of total charge $e$ and
minimal radius~\cite{Yuk49} is given by
$$
E = \frac 12 e^2 I(\kbar\mu^2).
$$
If we set this equal to the mass we find the equation
$$
e^2 \simeq - {8\pi \mu \over \log(\kbar\mu^2)}
$$
for the charge.

The energy density of a uniform, static, free scalar field is
given by
$$
T_{00} = \frac 12 \mu^2 \phi^2.
$$
The extra contribution due to vacuum fluctuations is 
$$
\vev{T_{00}}_0 = \frac 12 \mu^3 I(\kbar\mu^2).
$$
We have included an extra factor $\mu$ to account for the physical
dimensions of the field.  Considered just as a constant the vacuum
energy is not a very useful quantity unless somehow it can be
connected to the gravitational field equations. It is characteristic
of all vacuum-fluctuation calculations that the result is too large to
be a realistic source of a cosmological solution.  In some way `most'
of this very large constant must be subtracted. One way of doing this
is to consider the variation with respect to the space-time metric
just as the Casimir energy is calculated as that part of the vacuum
energy which depends on the distance.

Interpreted as a propagator on an ordinary manifold, $G$ would be
seen as associated to the non-local differential operator~\cite{PaiUhl50}
$$
\Delta_{NL} = e^{\kbar \bar \Delta/2} (\bar \Delta + \mu^2).
$$
This effective non-locality is due to the `quantization' of the
distance between the two points.  We have defended
elsewhere~\cite{MadMou98} the point of view~\cite{Des57, IshSalStr71,
CraSmo85, DopFreRob95, Oha97} that the regularization can be
considered in fact as being due to the gravitational field. To make
this point of view consistent with the results of the present section
one must consider the vacuum fluctuations as giving rise to a
microscopic field which disappears in the mean. In fact we shall argue
in Section~5 that flat space is to be considered as an idealized
limit.

There is a simple solid-state model for the space we have just
considered which has been used in the study of the fractional quantum
Hall effect. The $x$ and $y$ correspond to the cartesian components of
the guiding centers of the Landau orbits and the factor $e^{-\kbar
k^2/2}$ which arises here because of the effective non-locality acts
like the Debye-Waller factor.  We refer, for example, to
Meissner~\cite{Mei93} for further details.

It is straightforward to add a time coordinate and consider the
euclidean Laplace operator 
$$
\Delta = - \partial^2_t + \Delta_\kbar + \mu^2
$$
on the algebra $\c{A} = \c{C}(\b{R}) \otimes \c{A}_\kbar$
generated by the three hermitian elements $(t,x,y)$ and their
inverses. The differential calculus $\Omega^*(\c{A})$ is constructed by
adding to the 1-forms $dx$ and $dy$ the extra 1-form $dt$. The
density of euclidean vacuum action is given by
$$
I_E(\kbar\mu^2) = \frac{1}{(2\pi)^3\mu} 
\int {e^{-\kbar k^2/2} \over \omega^2 + k^2 + \mu^2} d\omega dk
= {1 \over \sqrt{32 \pi \kbar \mu^2}} e^{\kbar\mu^2/2} 
\Big(1 - \mbox{Erf}(\sqrt{\kbar\mu^2/2})\Big)
$$
where $\mbox{Erf}(x)$ is the error function. When 
$\kbar \mu^2 \to 0$
$$
I_E(\kbar\mu^2) = {1 \over \sqrt{32 \pi \kbar \mu^2}}
(1 - \sqrt{2\kbar\mu^2/\pi} + \cdots)
$$
and when $\kbar \mu^2 \to \infty$
$$
I_E(\kbar\mu^2) = \frac 1{4\pi \kbar\mu^2} + \cdots.
$$

\sect{The noncommutative Lobachevsky plane}

We shall define the noncommutative Lobachevsky plane to be the formal
$*$-algebra $\c{A}_h$ generated by two hermitian elements $x$ and $y$
which satisfy the commutation relation
\be
[x,y] = - 2 i h y                                               \label{2.1}
\ee
where $h \in \b{R}$ and the factor $-2$ is present for historical
reasons. We shall suppose that $h > 0$.  Both $x$ and $y$ are without
physical dimensions here. We define a differential calculus 
$(\Omega^*(\c{A}_h),d)$ over $\c{A}_h$ by
introducing~\cite{ChoMadPar98} a frame or Stehbein $\theta^a$ defined by
\be
\theta^1 = r y^{-1} dx, \qquad \theta^2 = ry^{-1} dy, \qquad    \label{theta}
\ee
where $r$ is a real parameter with the units of length.  The structure 
of the calculus is given by the commutation relations
\be
f \theta^a = \theta^a f, \qquad f \in \c{A}_h                  \label{2.2}
\ee
as well as the quadratic relations
\be
(\theta^1)^2 = 0, \qquad (\theta^2)^2 = 0, \qquad
\theta^1 \theta^2 + \theta^2 \theta^1 = 0.
\ee
More details of this have been given elsewhere~\cite{ChoMadPar98}.

We shall define~\cite{DubMadMasMou96} a metric $g$ as a bilinear map
\be
g(\theta^a \otimes \theta^b) = g^{ab}
\ee
where from (\ref{2.2}) the $g^{ab}$ must be real constants. We shall
choose $g^{ab} = \delta^{ab}$. From the structure relations
$$
d\theta^1 = - r^{-1} \theta^1 \theta^2, \qquad d\theta^2 = 0
$$
one concludes that the torsion-free metric connection has Gaussian
curvature $K$ given by $K = - r^{-2}$. 

The derivations $e_a$ dual to the 1-forms $\theta^a$ are defined by
$$
\begin{array}{ll}
e_1 x = r^{-1} y, &e_1 y = 0,\\[2pt]
e_2 x = 0, &e_2 y = - r^{-1} y.
\end{array}
$$
In terms of them the Laplace operator $\Delta_h$ can be 
written~\cite{Cho99} as
\be
- \Delta_h \phi = e_1^2 \phi + e_2^2 \phi + r^{-1} e_2 \phi, \label{Laplace}
\qquad \phi \in \c{A}_h.
\ee

First we recall the calculation of the propagator in the
commutative case. In the commutative limit $\Delta_h$ tends to the
ordinary Laplace operator on the Lobachevsky plane:
\be
\lim_{h\to 0} \Delta_h = \tilde \Delta = 
- r^{-2} \tilde y^2 (\partial_{\tilde x}^2 + \partial_{\tilde y}^2).
\ee
We have here introduced $(\tilde x, \tilde y)$ as the commutative limits
of the operators $(x,y)$. The spectrum of $\Delta_h$ in the commutative 
limit is given by~\cite{Ter85} the eigenvalue equation
\be
\tilde\Delta \phi (\tilde{x}, \tilde{y}) = 
\lambda_{k,\kappa} \phi (\tilde{x}, \tilde{y}).                  \label{3.2}
\ee                                                 
By the separation of variables 
$\phi(\tilde{x}, \tilde{y}) = f(\tilde{x})g(\tilde{y})$ we find the 
differential equations
\bea
&&\partial_{\tilde{x}}^2 f(\tilde{x}) = 
- k^2f(\tilde{x}),                                        \label{3.3}\\[4pt]
&&\tilde{y}^2 \partial_{\tilde{y}}^2 g(\tilde{y}) = 
(k^2 \tilde{y}^2 - r^2 \lambda_{k,\kappa}) g(\tilde{y}) \label{3.4}
\eea
where $k \in \b{R}$. If we define 
$\kappa^2 = r^2 \lambda_{k,\kappa} - 1/4$ then
$$
r^2 \lambda_{k,\kappa} = \kappa^2 + \frac 14 + r^2 \mu^2.
$$
The eigenvalues $\lambda_{k,\kappa}$ do not in fact depend on 
$k$ and are infinitely degenerate.  If we set then $z = ik\tilde{y}$ and
$g(\tilde{y}) = \sqrt{z}J(z)$, Equation~(\ref{3.4}) becomes the Bessel
equation
\be
J''(z) + \frac{1}{z}J'(z) + (1 + \frac{\kappa^2}{z^2})J(z) = 0.   \label{3.5}
\ee                                                      
A normalized set of eigenfunctions for the Laplace operator is given by
\be
\phi_{k,\kappa}(\tilde{x}, \tilde{y}) = 
e^{ik\tilde{x}} \pi^{-3/2}
\sqrt{\kappa \sinh \pi \kappa}  
\sqrt{\tilde{y}}K_{i\kappa}(|k|\tilde{y})                       \label{3.6}
\ee  
with $\kappa > 0$ and $k \neq 0$.  The case $\kappa < 0$ can be excluded
since
$$
K_{-\nu }(|k|\tilde{y}) = K_{\nu}(|k|\tilde{y}).
$$
The case $k = 0$ is also excluded since when $\tilde{y} \to 0$
\be
K_{i\kappa }(| k | \tilde{y}) \to \frac{1}{2}
\Gamma(i\kappa )\left(\frac{2}{|k| \tilde{y}}\right)^{i\kappa } + 
\frac{1}{2} \Gamma (-i\kappa )
\left(\frac{2}{|k| \tilde{y}}\right)^{-i\kappa}.              \label{expan}
\ee
If we set $\tilde x^i = (\tilde x, \tilde y)$ the completeness relation 
can be written as
\be
\delta^{(2)} (\tilde x^i - \tilde x^{i\prime}) = \int_{-\infty}^{+\infty} 
\int_0^\infty  \phi_{k,\kappa}(\tilde x, \tilde y)
\phi^*_{k,\kappa}(\tilde x^\prime, \tilde y^\prime) dk d\kappa
\ee
and the propagator is given by
\be
G (\tilde x^i, \tilde x^{i\prime}) = 
\int_{-\infty}^{+\infty} 
\int_0^\infty  {\phi_{k,\kappa}(\tilde x, \tilde y)
\phi^*_{k,\kappa}(\tilde x^\prime, \tilde y^\prime) 
\over \kappa^2 + \frac 14 + r^2 \mu^2} dk d\kappa.  
                                                                \label{Green}
\ee
The value of a tadpole diagram created by a source $J \to 0$
is given by the quantity
$$
I_L(\tilde x^i) = 
\lim_{\tilde x^{i\prime} \to \tilde x^i}G(\tilde x^i,\tilde x^{i\prime}).
$$
Because of the homogeneity of the space in fact $I_L$ cannot vary from
point to point; in ordinary field theory it is infinite.

Several interesting problems have been considered and
solved~\cite{ComHou85, Kub88, MleSte93, EmcNarThiSew94, Shc96, JanTel98}
on the Lobachevsky plane. In particular the spectrum of the Laplace
operator has been found~\cite{Ter85}. Recently~\cite{Cho99} moreover
the spectrum of the noncommutative operator (\ref{Laplace}) has been
calculated.

Consider now the noncommutative case.  It is to be noticed that although
the classical Lobachevsky plane is invariant under the reflection 
$\tilde x \to - \tilde x$ this is no longer the case when $h \neq 0$.
In the algebra $\c{A}_h$ any monomial $\phi(x,y)$ in $x$ and $y$ can be
factorized.  Therefore one can formally separate the variables in the
eigenvalue problem as before and the eigenvalue equation can be
decomposed into two differential equations. The equations for the factor
$f(x)$ are given by
\be
\begin{array}{l}
e_1^2 f(x) = - r^{-2} L_+^2y^2 f(x),                          \\[4pt]
e_1^2 f(x) = - r^{-2} L_-^2 f(x) y^2
\end{array}                                                 \label{3.7}
\ee
where $L_\pm \in \b{R}$.  Since the commutation relations $[y, e_2]$ and
$[\tilde{y}, \tilde y \partial_{\tilde{y}}]$ are of the same form, the
differential equation for $g(y)$ has the same form as that of
(\ref{3.4}) even though the algebra has changed:
$$
(e_2^2 + r^{-1} e_2) g(y) = 
r^{-2} (L_\pm^2 y^2 - \lambda_{k,\kappa}) g(y).
$$
Consider the function
$$
L(z) = {e^{z} - 1 \over z}\, k.
$$
It is related to the generating functional of the Bernoulli numbers,
which appears in one derivation of the general BaCH formula. For any
$k \in \b{R}$ let $e^{ikx}$ be defined as a formal power series in the
element $x$; formally $e^{ikx}$ is a unitary element of
$\c{A}_h$. Then from the action of $e_1$ on $x$ it follows that
\be
e_1 e^{ikx} = i r^{-1} L(2hk) y e^{ikx} 
= - i r^{-1} L(-2hk) e^{ikx} y.                                \label{3.9}
\ee                                                  
The solution of Equation~(\ref{3.7}) is given therefore by
\be
f(x) = e^{ikx},   \qquad L_\pm = \pm L(\pm 2hk).                 \label{3.10}
\ee                                                       

A family of formal solutions of the eigenvalue equation on the
quantum Lobachevsky plane which tend to normalized functions in the
commutative limit is given for $k \neq 0$, $\kappa >0$ by
\be
\phi_{k,\kappa}(x,y) = 
\pi^{-3/2} \sqrt{\kappa \sinh \pi \kappa}  
\sqrt{y}K_{i\kappa}(|L|y) e^{ikx}.                      \label{3.11}
\ee  
We have here introduced the quantity
$$
L = L_+(2hk).
$$
It plays the role of the linear momentum associated to $x$. The quantity
$L_-(2hk)$ is the linear momentum associated to $-x$. Although $|k|$
remains invariant under the map $k \to -k$ this is not the case for
$|L|$, a fact which is a manifestation of the breaking of parity by the
commutation relations. Because of the transposition rule
\be
e^{ikx} K(y) = K(e^{2hk}y) e^{ikx}                         \label{com-rel}
\ee
the expression for the eigenvectors can also be written with $y$ after
$x$. The 1-particle Hilbert space $\c{H}$ is the space
generated by the elements $\phi_{k,\kappa}(x,y)$.  The elements
$W$ and $G$ can be written then
\bea
&&W(x^\mu;x^{\nu\prime}) = \int_{-\infty}^{+\infty} 
\int_0^\infty  \phi_{k,\kappa}(x,y) \otimes
\phi^*_{k,\kappa}(x^\prime,y^\prime)
dk d\kappa, \nonumber\\ [6pt]
&&G(x^\mu;x^{\nu\prime}) = r^{-2}\int_{-\infty}^{+\infty} 
\int_0^\infty  \lambda_{k,\kappa}^{-1}
\phi_{k,\kappa}(x,y) \otimes
\phi^*_{k,\kappa}(x^\prime,y^\prime)
dk d\kappa.\nonumber
\eea

To proceed we must introduce a partial trace on the algebra $\c{A}_h$
which respects the $SL_h(2,\b{R})$ invariance.  This trace is a
complex-valued linear form on $\c{A}_h$ which is in some sense
translation invariant, and in the limit $h\to 0$ agrees with the
undeformed integral.  In the classical case, translation invariance is
equivalent to Stokes' theorem. Since we have an 
$SL_h(2,\b{R})$-invariant calculus, it is natural to define a trace of
an element of the algebra as the integral of the dual 1-form. For any 
$f \in \c{A}_h$ we set
$$
\tr(f) = \int f \theta^1 \theta^2
$$
where the volume 2-form, 
$$
\theta^1 \theta^2 = r^2 y^{-1} dx y^{-1} dy = r^2 y^{-2} dx dy,
$$
is invariant under the coaction of $ SL_h(2)$. We determine the
integral `over $x$' in turn by requiring that Stokes' theorem
\be
\int d\alpha =0.                                              \label{stokes}
\ee
hold for any 1-form $\alpha$. We write 
$\alpha = \alpha_x dx + \alpha_y dy$. In particular if $\alpha_x = 0$
and $\alpha_y = f(x) g(y)$ then from (\ref{stokes}) we find that
\be
\int df(x) g(y) y^{-2} dy = 
\int df(x) \int_0^\infty g(y) y^{-2} dy = 0.                  \label{zwei}
\ee
for any integrable function $g(y)$. To analyze this we notice that  
$(x+2ih) dx = dx x$ from which we deduce that
\bea
d(x^n) \squeeze &&= \Big((x+2ih)^{n-1} + x(x+2ih)^{n-2} + \dots + 
x^{n-1}\Big) dx \nonumber\\[4pt]
&&= x^{n-1}\Big(\sum_{k=0}^{n-1} (1 + 2ihx^{-1})\Big) dx
\nonumber\\[4pt]
&&= \frac{1}{2ih}x^n \Big((1+2ihx^{-1})^n-1\Big) dx
\nonumber\\[4pt]
&&= \frac{1}{2ih}\Big((x+2ih)^n-x^n\Big) dx.
\eea
We conclude that in general
\be
d f(x) = \frac 1{2ih} \Big(f(x+2ih)-f(x)\Big) dx,
\ee
which is a finite-difference operator.  Therefore
$$
\int \Big(f(x+2ih)-f(x)\Big) dx = 0.
$$
It follows then that 
$$
e^{-2hk} \int e^{ikx} dx = \int e^{ikx} dx
$$
and therefore it is consistent to set
\be
\tr_1(e^{ikx}) = 2\pi \delta(k).                               \label{tr-1}
\ee
Furthermore, in the representation given below we can identify 
\be
\tr_2 (f(y)) = \int_0^\infty f(y) y^{-2} dy.                   \label{tr-2}
\ee
Since $dy$
satisfies the commutation relation $ydy = dyy$ of an ordinary de~Rham
form on the undeformed Lobachevsky space we can suppose that
(\ref{tr-2}) holds in any case. 
The trace can be factorized then in the form
$$
\tr (e^{ikx} f(y)) = \tr_1 (e^{ikx}) \tr_2 (f(y)).
$$
and so, just as in the commutative case, we can set
$$
\tr (e^{ikx} f(y)) = 2 \pi \delta(k) \tr_2 (f(y)).
$$
If $x$ has a representation with a periodic spectrum then $k$ takes
discrete values and the right-hand side of this equation must be
replaced by $2\pi \delta_{k0}$. We note that for an arbitrary element
$f(x,y) \in \c{A}_h$ we have
$$
\tr (e^{ikx}f(x,y)) = \tr (f(x, e^{2hk} y)e^{ikx}).
$$
In general then 
$$
\tr (fg) \neq \tr (gf).
$$
The `trace' defines a state which is not a trace state.

Equations~(\ref{tr-2}) and (\ref{tr-1}) are all the properties of the
trace which we shall need.  Using them and the explicit expression
(\ref{3.11}) for the basis we find the orthogonality conditions
$$
\tr(\phi^*_{k,\kappa}(x,y) 
\phi_{k^\prime,\kappa^\prime}(x,y)) = 
\delta(k - k^\prime)
\delta(\kappa - \kappa^\prime).
$$

In order to use the general formalism we must first decide how to
introduce the annihilation and creation operators.  One possibility
for this is to introduce generators $\xi$ and $\eta$ which satisfy
the canonical commutation relations $[\xi,\eta] = 2ih$. One can then 
express $x$ and $y$ as
\be
x = \xi \eta - ih, \qquad y = \xi.                             \label{xi-y}
\ee
In the notation of (\ref{bar-delta}) this yields
\be
[\bar\xi, \bar\eta] = i h, \qquad
[\delta\xi, \delta\eta] = i h                                  \label{xi-eta}
\ee
and the condition (\ref{b-d}) is satisfied.  If we define
$$
\Lambda = e^{ix}, \qquad q = e^{-2h}
$$
we find the relation $y \Lambda = q \Lambda y$, which defines the quantum
space $\b{R}^1_q$. Because of the isotropy of the Lobachevsky plane the
Laplace operator is essentially reducible to that of a 1-dimensional
manifold. The extra dimension manifests itself as a difference in the
multiplicity of the eigenvalues. There is a certain formal analogy
between the solutions given here and the solutions~\cite{CerHinMadWes99}
to the Laplace operator in the quantum space $\b{R}^1_q$. 

If we express the eigenvectors in terms of the new generators we find
\be
\phi_{k,\kappa}(\xi,\eta)\ket{\bar p} = 
\pi^{-3/2} \sqrt{\kappa \sinh \pi \kappa}\,\sqrt{\xi}
K^*_{i\kappa}(|L| \xi)e^{ik(\xi\eta - i h)}\,\ket{\bar p}     \label{phi-1}
\ee
and therefore
\bea
&&\hskip -1cm \bra{\bar p^\prime}\phi_{k,\kappa}(x,y)\,
\phi^*_{k,\kappa}(x^\prime,y^\prime)\ket{\bar p} = \nonumber \\
[6pt] &&\pi^{-3} \kappa \sinh (\pi \kappa)
\bra{\bar p^\prime}K_{i\kappa}(|L|\xi) \sqrt{\xi} \;
e^{ik\xi\eta} e^{-ik\xi^\prime\eta^\prime}
\sqrt{\xi^\prime} K^*_{i\kappa}(|L|\xi^\prime) \ket{\bar p}.\nonumber 
\eea 
However as an added complication now the eigenvector is no longer
factorized as previously into a function of $\xi$ times a function of
$\eta$. Since we have supposed that $x$ and $x^\prime$ commute we can write 
$$
e^{ik\xi\eta} e^{-ik\xi^\prime\eta^\prime} =
e^{ik(\xi\eta - \xi^\prime\eta^\prime)} =
e^{-2ik(\bar\xi\delta\eta + \bar\eta\delta\xi)} = 
e^{-ik((\bar\eta - i \bar\xi) a + (\bar\eta + i \bar\xi)a^*)}.
$$
We have here introduced the annihilation operator $a$ and its adjoint
such that
$$
\delta \xi = \frac 12 (a + a^*), \qquad
\delta \eta = \frac 1{2i} (a - a^*), \qquad [a,a^*] = 2h.
$$
We cannot use the simple BaCH formula since
$$
[(\bar\eta - i \bar\xi)a,(\bar\eta + i \bar\xi)a^*] =
2h \Big(aa^* + (\bar\eta + i \bar\xi)(\bar\eta - i \bar\xi)\Big)
$$
does not commute with $(\bar\eta - i \bar\xi) a$ and 
$(\bar\eta + i\bar\xi) a^*$. In fact these three operators form a
basis of the Lie algebra of $SL(2,\b{C})$. The 
$(\bar\eta + i\bar\xi)$ is essentially the extra annihilation operator
$b$ introduced in the previous section and we have thus a tensor
product of two harmonic-oscillator representations.  It would seem
that the propagator is impossible to calculate using the
decomposition (\ref{xi-eta}). 

If we use $x$ and $y$ as generators and follow the prescription of
Section~2 we find that using an ordinary tensor product
\be
[\bar x, \bar y] = - i h \bar y, \qquad 
[\delta x, \delta y] = - i h \bar y                           \label{ddelta}
\ee
as in the previous section but the condition (\ref{b-d}) is not satisfied:
\be
\begin{array}{lll}
[\bar x, \delta x] = 0, &\quad
&[\bar x, \delta y] = - i h \delta y, \\[4pt]
[\bar y, \delta x] = i h \delta y, &\quad
&[\bar y, \delta y] = 0.
\end{array}                                                   \label{delta}
\ee
This means that $\bar x$ acts on $\c{F}$ as well as $\c{D}$ in the
product (\ref{factor}). This point can be improved upon by a change of
generators. First we note that the algebra generated by
$(\bar x, \bar y)$ can be identified with the algebra $\c{A}_h$ and
that the differential calculus $(\bar\Omega^*(\c{A}_h),\bar d)$ defined
by the relations (\ref{delta}),
$$
\begin{array}{lll}
[\bar x, \bar d \bar x] = 0, &\quad
&[\bar x, \bar d \bar y] = - i h \bar d \bar y, \\[4pt]
[\bar y, \bar d \bar x] = i h \bar d \bar y, &\quad
&[\bar y, \bar d \bar y] = 0
\end{array}
$$
is the same as the original $(\Omega^*(\c{A}_h),d)$. In fact
one finds that the frame $(\bar\theta^1, \bar\theta^2)$ defined
by
$$
\bar\theta^1 = r\bar d \bar x - r\bar x\bar y^{-1} \bar d \bar y, \qquad
\bar\theta^2 = r\bar y^{-1} \bar d \bar y
$$
satisfies the same relations as the frame (\ref{theta}). In the
commutative limit the new frame is the old one expressed in the new
coordinates given by the involution 
$$
\tilde{\bar x} = \phi(\tilde x) = \tilde x \tilde y^{-1}, \qquad
\tilde{\bar y} = \phi(\tilde y) = \tilde y^{-1}.
$$
General covariance would seem to suggest then that one introduce an
annihilation operator such that
\be
\delta x = \frac 12  (a + a^*) + \frac 1{2i} \bar x (a - a^*), \qquad
\delta y = \frac 1{2i} \bar y(a - a^*), \qquad [a,a^*] = h.    \label{y}
\ee
If one did this one would find Equations~(\ref{delta}) to be
equivalent to the conditions 
$$
[\bar x, a] = 0, \qquad [\bar y, a] = 0
$$
but that the second of the commutation relations (\ref{ddelta}) cannot
be satisfied. This is to be expected since differential forms
naturally satisfy anticommutation relations. The expressions (\ref{y})
come from the identification
$$
\theta^1 = \frac 12 (a + a^*), \qquad 
\theta^2 = \frac 1{2i} (a - a^*)
$$
and the relations satisfied by the frame would imply the relations 
$a^2 = 0$, $[a,a^*]_+ =0$.

One can express $\phi$ as the commutative limit of the change to new
generators given by
$$
\bar x^\prime = x y^{-1} - i h, \qquad \bar y^\prime = y^{-1}.
$$
This transformation is closely related to that given by
(\ref{xi-y}). Under the change of parameter $h \to -2h$ we can identify
$\bar x^\prime = \eta$ and $\bar y^\prime = \xi^{-1}$. The 
$(\bar x^\prime$, $\bar y^\prime)$ satisfy the same commutation 
relation as the $(\bar x$, $\bar y)$ except for a change in sign of $h$.
We have here defined the differential calculus directly in terms of
the algebra; in particular, we have deduced the module structure of
the 1-forms from the commutation relation.  This was possible since
both the algebra and the differential calculus are defined in terms of
the same $R$-matrix~\cite{Agh93}.

A more promising decomposition uses the generator $w$ formally defined
by the equation $y = e^{-w}$. Using it the commutation relation
(\ref{2.1}) becomes
$$
[x,w] = 2ih
$$
and in the commutative limit $r \tilde w$ is the geodesic distance along 
the $\tilde y$-axis. Following the prescription of
Section~2 we find that, using an ordinary tensor product,
$$
[\bar x, \bar w] =  i h , \qquad 
[\delta x, \delta w] = i h 
$$
and that the condition (\ref{b-d}) is now satisfied. 

The Equation~(\ref{phi-1}) becomes
\bea
&&\phi^*_{k,\kappa}(x^\prime,y^\prime)\ket{\bar p} = \nonumber
\\ [6pt] &&\hskip 2cm \pi^{-3/2} e^{-ik\bar x}
e^{-ik\delta x} \sqrt{\kappa \sinh \pi \kappa}\sqrt{y^\prime}
K^*_{i\kappa}(|L|y^\prime)\,\ket{\bar p}\nonumber
\eea
and therefore
\bea
&&\hskip -1cm \bra{\bar p^\prime}\phi_{k,\kappa}(x,y)\,
\phi^*_{k,\kappa}(x^\prime,y^\prime)\ket{\bar p} = \nonumber \\[6pt] 
&&\hskip 1cm \pi^{-3} \kappa \sinh (\pi \kappa)
\bra{\bar p^\prime}K_{i\kappa}(|L|y) 
\sqrt{y} e^{-2ik\delta x} \sqrt{y^\prime}
K^*_{i\kappa}(|L|y^\prime) \ket{\bar p}.\nonumber 
\eea 
As above we introduce the annihilation operator $a$ and its adjoint
such that
$$
\delta x = \frac 12 (a + a^*), \qquad
\delta w = \frac 1{2i} (a - a^*), \qquad [a,a^*] = 2h.
$$
Using again the BaCH formula we find that
\bea 
&&\hskip -1cm\bra{\bar p^\prime}\phi_{k,\kappa}(x,y)\,
\phi^*_{k,\kappa}(x^\prime,y^\prime)\ket{\bar p} = \nonumber \\
[6pt] &&\hskip 1cm \pi^{-3} \kappa \sinh (\pi \kappa) 
e^{- h k^2} \bra{\bar p^\prime} K_{i\kappa}(|L|y)
\sqrt{y} e^{-ika^*} e^{-ika} \sqrt{y^\prime}
K^*_{i\kappa}(|L|y^\prime)\ket{\bar p}.\nonumber 
\eea 
Using the transposition rules
$$
w e^{ika^*} = e^{ika^*} (w - hk), \qquad
e^{ika}w^\prime = (w^\prime - hk) e^{ika}
$$
we conclude therefore that
\bea
&&\hskip -0.5cm \bra{\bar p} 
G(x,y;x^\prime,y^\prime)\ket{\bar p} = \nonumber \\ [6pt]
&&\pi^{-3}\int_{-\infty}^{+\infty} 
\int_0^\infty {\kappa \sinh (\pi \kappa) e^{-hk^2} \over
\kappa^2 + \frac 14 + r^2 \mu^2}\;
\bra{\bar p}K_{i\kappa}(|L| e^{-hk} y) \times\nonumber\\[6pt]
&&\hskip 1cm  e^{-hk} e^{-\bar w}
K^*_{i\kappa}(|L| e^{-hk} y^\prime)
\ket{\bar p} d\kappa dk.\nonumber
\eea
This can be expressed as an integral over positive values of $k$: 
\bea
&&\hskip -0.5cm \bra{\bar p} 
G(x,y;x^\prime,y^\prime)\ket{\bar p} = \nonumber \\ [6pt]
&&2\pi^{-3} \int_0^{+\infty} 
\int_0^\infty {\kappa \sinh (\pi \kappa) e^{-hk^2} \cosh(hk) \over
\kappa^2 + \frac 14 + r^2 \mu^2}\times\nonumber\\[6pt]
&&\hskip 1cm \bra{\bar p}K_{i\kappa}(h^{-1}\sinh(hk)y) 
e^{-\bar w} K^*_{i\kappa}(h^{-1}\sinh(hk) y^\prime)
\ket{\bar p} d\kappa dk.\nonumber
\eea
The integral can be simplified by introducing the integration
variable
$$
h l = \sinh(hk) e^{-\bar w}.
$$
It becomes then
$$
\bra{\bar p} G(x,y;x^\prime,y^\prime) \ket{\bar p} = 
2 \pi^{-3} \int_0^\infty\int_0^\infty 
{\kappa \sinh (\pi \kappa)\over \kappa^2 + \frac 14 + r^2 \mu^2}\, 
F(\kappa,l)d\kappa dl.
$$
where
\be
F(\kappa,l) =
\bra{\bar p,0}K_{i\kappa}(l e^{+\delta w})
e^{-h^{-1}\mbox{\scriptsize{arcsinh}}^2(hle^{\bar w})}  
K^*_{i\kappa}(l e^{-\delta w}) \ket{\bar p,0}.
\ee
This function is not manifestly independent of the state $p$, that is,
of the value of $\bar w$. We can write 
$$
F(\kappa,l) = G(l)\, H(\kappa,l)
$$
where
$$
H(\kappa,l) = {}_F\bra{0}K_{i\kappa}(l e^{+\delta w})
K^*_{i\kappa}(l e^{-\delta w}) \ket{0}_F.
$$
is manifestly independent of $\bar w$ but
\be
G(l) = {1 \over {}_D\vev{\bar p\,|\,\bar p}_D}
{}_D\bra{\bar p} e^{-h^{-1}\mbox{\scriptsize{arcsinh}}^2(hle^{\bar w})} 
\ket{\bar p}_D                                                  \label{G}
\ee
is not.

In an attempt to clarify this we consider an explicit representation
of the algebra.  On the Hilbert space $L^2(\b{R},d\alpha)$ one has the
representation given on smooth functions by
$$
(\bar x f)(\alpha) = i h \partial_\alpha f(\alpha), \qquad
(\bar w f)(\alpha) = \alpha f(\alpha).
$$
A convenient basis is given by $\ket{p} = e^{ip\alpha/h}$. We find
then the expression
$$
e^{\bar w}\ket{\bar p} = e^{\alpha} \ket{\bar p}
$$
and the function (\ref{G}) can be written as
$$
G(l) = {1 \over {}_D\vev{\bar p\,|\,\bar p}_D}{}_D\bra{\bar p}
e^{-h^{-1}\mbox{\scriptsize{arcsinh}}^2(hle^{\bar w})} \ket{p}_D = 
\lim_{\alpha\to\infty}{1 \over 2\alpha_0}\int_{-\alpha_0}^{+\alpha_0} 
e^{-h^{-1}\mbox{\scriptsize{arcsinh}}^2(hle^\alpha)} d\alpha = \frac 12.
$$
This is certainly independent of $\alpha$ but depends in the choice of
basis; the states $\ket{p}$ are plane-wave states and the $\bar w$
`coordinate' is `smeared out' over the entire line. Another choice of 
representation is obtained by interchanging $\bar x$ and $\bar w$. 
That is, with
$$
(\bar x f)(\alpha) = \alpha f(\alpha), \qquad
(\bar w f)(\alpha) = i h \partial_\alpha f(\alpha).
$$

In this representation $\bar w$ is diagonal and $p$ is an eigenvalue,
a measure of the geodesic distance along the $\bar y$-axis. It leads
to
$$
G(l) = e^{-h^{-1}\mbox{\scriptsize{arcsinh}}^2(hle^p)},
$$
which definitely depends on $p$. Because of the discussion that led to 
Equation~(\ref{y}) we shall argue below that the results are only
valid at the point $p = 0$ on the $\bar y$-axis. This would imply that
$$
G(l) = e^{-h^{-1}\mbox{\scriptsize{arcsinh}}^2(hl)}.
$$
We found in the previous case that the result depended on our choice
of representation of the $\delta q^\mu$-algebra; we find here that it
depends also on the representation of the $\bar q^\mu$.

We set $\kbar = 2 h r^2$ and we define as previously $I_L(\kbar\mu^2)$ 
by the equation 
$$
\bra{\bar p} G(x,y;x^\prime,y^\prime)\ket{\bar p} =
I_L (\kbar\mu^2) \vev{\bar p\,|\, \bar p}.
$$
We have then 
\be
I_L (\kbar\mu^2) = 2 \pi^{-3}
\int_0^\infty\int_0^\infty {\kappa \sinh (\pi \kappa) \over
\kappa^2 + \frac 14 + r^2 \mu^2}\; G(l) H(\kappa, l) d\kappa dl. \label{I-L}
\ee
We shall leave the evaluation of $H$ to a future publication. The
integral $I_L(\kbar\mu^2)$ can be estimated however to leading order
from the fact that the uncertainty relations, as encoded in the
commutation relation between $a$ and its adjoint, imply that
${}_F\bra{0}\delta w\ket{0}_F \gtrsim h$. From this we can deduce
that
$$
I_L(\kbar\mu^2) \simeq - \frac 1{4\pi} \log (\kbar \mu^2) + \cdots.
$$
All the interesting information is in the sub-dominant terms, 
which appear in the difference
$$
\Delta \vev{T_{00}}_0 = \frac 12 \mu^3 (I_L (\kbar\mu^2) - I(\kbar\mu^2)),
$$
in the energy density with and without the curvature. 

On the fuzzy sphere~\cite{Mad99} of radius $r$ the laplacian has $n$
distinct eigenvectors $f_s$ with associated eigenvalues 
$\omega_s^2 = s(s+1)r^{-2}$ of multiplicity $2s+1$. Let $\{\ket{i}\}$
be a basis of coherent states and define $I_S(\kbar\mu^2)$ by the
equation
$$
\bra{i,i} G(x^a;x^{a\prime}) \ket{i,i} = 
4 \pi r^2 I_S(\kbar\mu^2)\,\vev{i\,|\,i}^2.
$$
Because of the properties of coherent states $I_S(\kbar\mu^2)$ will 
be independent of the state. As in the case of the plane we write
$f_s(x^a) = f_s(\bar x^a - \delta x^a)$ and 
$f_s(x^{a\prime}) = f_s(\bar x^a + \delta x^a)$. If $\ket{i}$ is the state
concentrated on the north pole of the sphere then for large $n$ we can
write the commutation relations as $([x,y] - \kbar) \ket{i} = 0$ and
identify the sphere with the tangent plane. Comparing the two cases
one finds that for large $n$
\be
I_S(\kbar\mu^2) \simeq
\frac 1{8\pi} \sum_{s=0}^{n-1} {2s+1 \over s(s+1) + r^2 \mu^2} =
\frac 1{4\pi} \sum_{s=1/2}^{n-1/2} {s \over s^2 - \frac 14 + r^2 \mu^2}
                                                                   \label{I-S}
\ee
from which we deduce that
$$
I_S(\kbar \mu^2) \simeq I\Big(\kbar(\mu^2 - \frac 1{4r^2})\Big).
$$
We have here used the relation $4\pi r^2 \simeq 2\pi\kbar\, n$ between
the area of the sphere and the area of the fundamental cell. We find
therefore, when $\kbar\mu^2 \to 0$ and $r\mu \to\infty$, that
$$
I_S(\kbar\mu^2) - I(\kbar\mu^2) \sim \frac 1{32\pi r^2\mu^2}.
$$
It is tempting to use the difference  
$$
\Delta \vev{T_{00}}_0 \sim \frac 1{64\pi} \frac \mu{r^2}
$$
of $\vev{T_{00}}_0$ as a source in the gravitational field equations.
We shall return to a similar calculation in dimension 4 below.

If we compare (\ref{I-L}) with (\ref{I-S}) we see that the eigenvalues
are identical except for a change in sign in the curvature term. We
can therefore reasonably suppose that
$$
I_L(\kbar \mu^2) \simeq I\Big(\kbar(\mu^2 +\frac 1{4r^2})\Big).
$$

If we define $z = x + iy$ then the commutation relation (\ref{2.1})
which define the algebra $\c{A}_h$ can be written in the form
$$
[z, \bar z] = 2ih (z - \bar z).
$$
There is a Cayley transformation
$$
z^\prime = {z - i \over z + i}
$$
from the Lobachevsky plane to the Poincar\'e disc. To compare the
calculations of this section with those of the flat plane one might
think that it would be simpler to use the disc but the commutation
relation in terms of $z^\prime$ is rather complicated:
\be
[z^\prime, \bar z^\prime] = 
\frac 14 (1 - \bar z^\prime) (1- z^\prime) 
[z, \bar z] (1-z^\prime)(1-\bar z^\prime) = 
- h( 1-z^\prime)(1- \bar z^\prime z^\prime)
( 1 - \bar z^\prime) + o(h^2).                            \label{ud1}
\ee
One definite advantage of the disc is that in the commutative limit
there is one point $\tilde z^\prime = 0$, at which the metric assumes
the gaussian normal form, with the first derivatives of the components
equal to zero. The Poincar\'e disc has also been `quantized' by
Berezin~\cite{Ber75}, with the commutation relation
\be
[z^\prime, \bar z^\prime] = 
h (1 - z^\prime \bar z^\prime)(1 - \bar z^\prime z^\prime).   \label{ud2}
\ee
There is no obvious relation between the two commutation
relations~(\ref{ud1}) and~(\ref{ud2}).

It would seem that when studying a noncommutative version of a general
manifold one first has to choose a system of coordinates which are
gaussian normal at a point $\tilde q^\mu = \tilde q^\mu_0$. The 
corresponding generators of the noncommutative tensor-product algebra 
must be then studied in a coherent state $\ket{0}$ with 
$\bra{0}\bar q^\mu \ket{0} = \tilde q^\mu_0$. The propagator in
this state is the noncommutative version of the propagator at the
point $\tilde q^\mu = \tilde q^\mu_0$. From the above experience with
the Lobachevsky plane we conclude that even in the case of a
noncommutative version of a homogeneous manifold $H$, with therefore
$I_H(\kbar\mu^2)$ a constant function, the propagator can only be
calculated in a state which is localized about a classical point at
which the metric is at least euclidean, if not gaussian normal. The
problem is due to the fact that even in the simplest of noncommutative
geometries the relation of the noncommutative structure to the metric
is not well understood. This is already apparent at the commutative
limit. The Poisson structure defined in this limit is in the canonical
form in a system of coordinates which in general has no obvious
preferred relation to the metric.

With the addition of an extra time coordinate the algebra
becomes $\c{A} = \c{C}(\b{R}) \otimes \c{A}_h$ generated by the
three hermitian elements $(t,x,y)$ and their inverses. The differential
calculus $\Omega^*(\c{A})$ is constructed by adding to the two `space'
1-forms $\theta^a$ the time 1-form $\theta^0 = dt$ and imposing the
standard relations.  In the limit $h \to 0$, $\c{A}$ becomes an algebra
of time-dependent functions on the Lobachevsky plane and
$\Omega^*(\c{A})$ the corresponding de~Rham differential calculus. 
The euclidean Laplace operator of a free scalar field is
$$
\Delta = - \partial^2_t + \Delta_h + \mu^2.
$$
We find then the curved variant
$$
I_{EL} (hr^2\mu^2) = 2 \pi^{-3}r^2\mu^{-1}
\int_{-\infty}^{+\infty} \int_0^\infty
\int_0^\infty {\kappa \sinh (\pi \kappa)  \over
r^2 \omega^2 + \kappa^2 + \frac 14 + r^2 \mu^2}\; 
G(l) H(\kappa,l) d\kappa dl d\omega
$$
of $I_E(\kbar\mu^2)$. It is also possible to let $h$ vary with time but
because the curvature does not depend on the value of $h$ there can be
no dynamical evolution.  Since the space is completely isotropic and
homogeneous one might speculate that there is variation in $h$ (in
space and time) only in the presence of inhomogeneities and that these
latter relax to yield a homogeneous space and a constant $h$. One
would have to consider the time evolution of perturbations of the
Lobachevsky metric to determine whether or not this is the case.

The cut-off effect which we have found was obtained using an ordinary
tensor product. As in the flat case, and for the same reasons, one
finds that the use of a braided tensor product will yield a 
propagator which is independent of $h$ and which can be identified with
the divergent propagator of the commutative limit~\cite{MadSte99}.

\sect{The noncommutative flat 4-space}

We define the noncommutative flat 4-space as the algebra $\c{A}_\kbar$
generated by four elements $q^\mu = x^\mu$ which satisfy the
commutation relations~\cite{DopFreRob95}
$$
[x^\mu,x^\nu] = i \kbar J^{\mu\nu}
$$
where $J^{\mu\nu}$ is a non-degenerate matrix of real numbers.  The
associated differential calculus $\Omega^*(\c{A}_\kbar)$ is defined by
the relations $[x^\mu, dx^\nu] = 0$. If we introduce the derivations
$$
e_\alpha = \ad \lambda_\alpha, \qquad 
\lambda_\alpha = {1 \over i \kbar} J^{-1}_{\alpha\mu} x^\mu
$$
dual to $dx^\mu$ then an appropriate generalization~\cite{MadMou98} of
the Laplace operator $\Delta$ with mass $\mu$ is given by
$$
\Delta = \Delta_\kbar + \mu^2 = - \sum_\alpha e^2_\alpha + \mu^2.
$$
For each $k \in \b{R}$ we introduce the elements 
$u_\mu(k) \in \c{A}_\kbar$ defined by
$$
u_\mu (k) = e^{ik x^\mu}.
$$
They satisfy the commutation relations 
$$
u_\mu(k_1)u_\nu(k_2) = 
q^{J^{\mu\nu} k_1 k_2\kbar} u_\nu(k_2)u_\mu(k_1), \qquad 
q = e^{-i}.
$$
A basis for the Hilbert space $\c{H}$ is given by the eigenvectors
$$
\phi_{k} = u_1(k_1)u_2(k_2)u_3(k_3)u_4(k_4) =
e^{ik_1x^1}e^{ik_2x^2}e^{ik_3x^3}e^{ik_4x^4}, \qquad k = (k_1,k_2,k_3,k_4)
$$
of $\Delta$.  The corresponding eigenvalues are 
$\lambda_k = k^2 + \mu^2$ where we have set 
$k^2 = g^{\mu\nu} k_\mu k_\nu$. The element $G$ can be written then
$$
G(x^\mu;x^{\mu\prime}) =
\frac{1}{(2\pi)^4} \int (k^2 + \mu^2)^{-1}
\phi^{\phantom{\prime *}}_k \otimes 
\phi^*_k \, dk,\qquad dk = dk_1dk_2dk_3k_4.
$$

We must introduce a partial trace on $\c{A}_\kbar$. This can be done only
through a representation.  The only properties which we shall need are
the identities
$$
\tr(u^*_\mu(k^\prime)u^{\phantom{*}}_\nu(k)) = 
2 \pi \delta(k^\prime - k) g_{\mu\nu}.
$$
That is:
$$
\tr(\phi^*_{k^\prime} \phi^{\phantom{*}}_k) = 
(2\pi)^4\delta^{(4)}(k^\prime - k).
$$
It is most convenient to choose a generalization of the second
representation given in Section~3, the one which is reducible and
non-singular in the limit $\kbar \to 0$. We represent $\c{A}_\kbar$ as
an algebra of operators on $L^2(\b{R}^4,dx)$ defined  on 
$f(\alpha^\lambda)\in L^2(\b{R}^4,dx)$ by
$$
u_\mu(k)f(\alpha^\lambda) = 
e^{ik\alpha^\mu} f(\alpha^\lambda + \frac 12 \kbar J^{\lambda\mu} k).
$$
A convenient basis for
$L^2(\b{R}^4,dx)$ is $\ket{p} = e^{ip_\lambda \alpha^\lambda}$ 
with $p_\lambda\in \b{R}$. We have then
$$
u_\mu(k)\ket{p} = 
q^{\frac 12 \kbar J^{\mu\nu} k p_\nu}\ket{p_1+k\delta_{\mu 1},
p_2+k\delta_{\mu 2},p_3+k\delta_{\mu 3},p_4+k\delta_{\mu 4}}.
$$
The eigenvectors $\phi_k$ have matrix elements defined by
$$
\phi_k\ket{p} = 
q^{\frac 12 \kbar J^{\mu\nu} k_\mu p_\nu}\ket{p + k}.
$$

The commutation relations (\ref{diff-rel}) become in this case
$$
[\bar x^\mu, \bar x^\nu] = \frac 12 i \kbar J^{\mu\nu},\qquad
[\delta x^\mu, \delta x^\nu] = \frac 12 i \kbar J^{\mu\nu}.
$$
We introduce the operators $a_1$ and $a_2$ as previously in Section~2 
and we write
\be
\delta x^\mu = J^{\mu}_1 a_1 + J^{\mu*}_1 a^*_1 + 
J^{\mu}_2 a_2 + J^{\mu*}_2 a^*_2                         \label{modes}
\ee
from which we conclude that
$$
J^{[\mu}_1 J^{\nu]*}_1 + J^{[\mu}_2 J^{\nu]*}_2
= \frac 12 i J^{\mu\nu}.
$$
We have therefore in the basis 
$\ket{\bar p,k} \equiv \ket{\bar p}_D \otimes \ket{k}_F$, with as
before $\ket{\bar p} \equiv \ket{\bar p, 0}$,
\bea
u_\mu^*(k)\ket{\bar p} \squeeze&&=
e^{-ik x^\mu}\ket{\bar p} = 
e^{-ik(\bar x^\mu - \delta x^\mu)}
\ket{\bar p} \nonumber\\[6pt]
&&= e^{-ik\bar x^\mu} e^{ik \delta x^\mu}
\ket{\bar p}\nonumber\\[6pt]
&&= e^{-ik\bar x^\mu} e^{ik(J^{\mu}_1 a + J^{\mu*}_1 a^*)}
e^{ik(J^{\mu}_2 b + J^{\mu*}_2 b^*)} \ket{\bar p}.\nonumber
\eea
Using the BaCH formula we find that
$$
u_1^*(k_1)\ket{\bar p} =
e^{-ik_1\bar x^1}e^{-\kbar(|J_1^1|^2 + |J_2^1|^2) k_1^2/2} 
e^{ik_1(J^{1*}_1 a^* + J^{1*}_2 b^*)} \ket{\bar p}
$$
and therefore
$$
\phi^*_k \ket{\bar p} = e^{-ik_\mu\bar x^\mu}
e^{-\kbar K^{\mu\nu} k_\mu k_\nu/4} e^{i\omega(k_\mu)}
e^{ik_\mu(J^{\mu*}_1 a^* + J^{\mu*}_2 b^*)}\ket{\bar p}.
$$
The $\omega$ is an unimportant phase factor and we have introduced the
diagonal tensor
$$
\frac 12 K^{\mu\nu} =  \mbox{diag}(|J_1^1|^2 + |J_2^1|^2,\; 
|J_1^2|^2 + |J_2^2|^2,\; |J_1^3|^2 + |J_2^3|^2,\; |J_1^4|^2 + |J_2^4|^2).
$$
The expectation value of the propagator is given by the expression
$$
\bra{\bar p}G(x^\mu;x^\nu)\ket{\bar p} =
\mu^2 I(\kbar\mu^2,K) \vev{\bar p\,|\, \bar p}
$$
with
$$
I(\kbar\mu^2,K) =
\frac{1}{(2\pi)^4\mu^2} \int {e^{-\kbar K^{\mu\nu}k_\mu k_\nu} \over 
k^2 + \mu^2} dk.
$$

We must now address the delicate question of (euclidean) Lorentz
invariance. There are two attitudes one can take. One can suppose that
Lorentz invariance is exact at all scales. One must then
add~\cite{DopFreRob95} the $J^{\mu\nu}$ as six extra coordinates, minus
possibly two because of two invariants which can be formed. Either one
considers that there is no momentum associated to these coordinates, in
which case one can take an average value over them and the problem is
solved as above, or one can consider them to be ordinary coordinates
like the four visible ones, in which case they would have to be
`quantized' also. If this be so the $J^{\mu\nu}$ cannot lie in the
center~\cite{Sny47a}.  Alternatively one can admit that the tensor
$J^{\mu\nu}$ breaks Lorentz invariance on the scale of $\kbar$.  This
manifests itself by the existence of the vectors $J_1^\mu$ and
$J_2^\mu$. However there is also an ambiguity in the choice of creation
and annihilation operators, described by the symplectic group here of
dimension 10. It is always possible then to choose $a_1$ and $a_2$ so
that 
$$
K^{\mu\nu} = g^{\mu\nu}.
$$
We shall suppose that this has been done.  The issue of Lorentz
invariance will not appear explicitly then, except to the extent that
our calculations are not invariant under the symplectomorphism group.
This is fortunate since we have motivated the introduction of
noncommuting coordinates by the desire to maintain Lorentz invariance.

The integral $I(\kbar\mu^2) = I(\kbar\mu^2,g)$ is given by
$$
I(\kbar\mu^2) =
\frac{1}{(2\pi)^4\mu^2} \int
{e^{-\kbar k^2} \over k^2 + \mu^2} dk = 
\frac{1}{16\pi^2} \Big(\frac 1{\kbar \mu^2} + 
e^{\kbar\mu^2} \mbox{Ei} (-\kbar\mu^2)\Big).
$$
For all values of $\kbar\mu^2$ the function $I(\kbar\mu^2)$
is concave. When $\kbar\mu^2 \to 0$
\be
I(\kbar\mu^2) = \frac{1}{16\pi^2}
\Big(\frac 1{\kbar\mu^2} + \log (\kbar \mu^2) + \cdots\Big)      \label{I-K}
\ee
and when $\kbar\mu^2 \to \infty$
$$
I(\kbar\mu^2) = \frac{1}{16\pi^2(\kbar\mu^2)^2} 
\Big(1 - \frac 2{\kbar\mu^2} + \cdots\Big).
$$
We would like eventually to compare the expression (\ref{I-K}) with a
curved-space analogue in the limit of vanishing curvature; this would
supply a preferred Lorentz frame in the limit.

The density of (euclidean) action of a uniform, static, free scalar 
field is given by
$$
\Gamma = \frac 12 \mu^2 \int \phi^2.
$$
The quantity 
$$
\vev{\Lambda}_0 = \frac 12 \kbar \mu^4 I(\kbar\mu^2)          
$$
can be interpreted~\cite{Wei89} as a contribution of the 
scalar-field vacuum fluctuations to the cosmological constant.  We would 
like to be able to compare this with a noncommutative `curved-space'
configuration but in dimension 4 we have thus far only been able to
consider the flat case. In the absence of any other information we
suppose the dominant contribution to be obtained by a substitution of
the form
\be
I_K(\kbar \mu^2) \simeq 
I\Big(\kbar(\mu^2 + \alpha K )\Big)                  \label{Ansatz}
\ee
where $\alpha$ is a constant and $K$ is some local mean curvature. We 
find then from (\ref{I-K}) that
\be
\Delta\vev{\Lambda}_0 \simeq - \frac 1{32\pi^2} \alpha K 
\Big(1 - {\alpha K \over \mu^2}\Big).                          \label{cc}
\ee
If the space-time has constant curvature $K$ then 
$\Delta\vev{\Lambda}_0 = - 3 K$. Consistency requires then that 
$$
\alpha =  \frac{32}3 \pi^2 + o(K\mu^{-2}).
$$

If one is interested in a cosmological solution of type
Friedmann-Robertson-Walker then one can identify 
$$
\kbar \to 8\pi G_N, \qquad K \to a(t)^{-2}
$$
and Wick-rotate to real time. One must also replace (\ref{cc}) by
\be
\kbar \vev{\rho}_0 \simeq \frac 1{32\pi^2} {\alpha \over a^2} 
\Big(1 - {\alpha \over \mu^2 a^2}\Big), \qquad 
\vev{\rho}_0 \equiv \Delta\vev{T_{00}}_0                       \label{T00}
\ee
since a solution cannot be found with a varying effective ccosmological
constant. One obtains in the flat case $(k=0)$ the equation
$$
\dot a^2 = {\alpha\over 96\pi^2} \Big(1 - {\alpha \over \mu^2 a^2}\Big)
$$
which has a bounce solution given by
$$
a(t) = \sqrt{{\alpha\over \mu^2}
\left(1 + {\mu^2 t^2 \over 96\pi^2}\right)}\,.
$$
The effective pressure is negative~\cite{WanCalOstSte99}:
$$
\kbar \vev{p}_0 \simeq - {1\over 96\pi^2} {\alpha\over a^2}
\Big(1 + {\alpha \over \mu^2 a^2}\Big)
$$
and the strong energy condition is violated:
$$
\kbar (\vev{\rho}_0 + 3 \vev{p}_0) \simeq
- {1 \over 16 \pi^2} {\alpha^2 \over \mu^2 a^4} < 0.
$$

The energy which is the effective source of the solution is the
difference between two vacuum energies and its sign depends simply on
which of the two is the larger. The minimal radius is given by 
$\mu^2 a^2(0) = \alpha$ and at this value of $t$ the approximation to
$I_K(\kbar\mu^2)$ which we have used is no longer valid. This is
evident from the fact that the expression (\ref{T00}) for $\vev{\rho}_0$
vanishes at the bounce and it should be maximum there.

Another problem which one can consider is the `self-consistent' mass
calculation~\cite{NamJon61} based on a 1-loop approximation to the
Schwinger-Dyson equation. With an interaction of the form
$\lambda\phi^4$ a scalar field acquires a mass $\mu$ which must
satisfy the equation $\mu^2 = \lambda \mu^2 I(\kbar\mu^2)$.  That is,
to leading order
$$
\lambda \sim 8\pi^2\kbar\mu^2.
$$
If $\kbar$ is identified with the square of the Planck length this
would imply an interaction constant $\lambda$ slightly larger than
$10^{-20}$. If on the other hand we require that $\lambda \sim 1$
then this would imply that $\kbar\mu^2 \sim 1/(8\pi^2)$.

The noncommutative torus is the formal algebra generated by the
$u_\mu$ for arbitrary fixed values of the $k_\mu$. It was the first
noncommutative geometry on which a Yang-Mills action was
proposed~\cite{ConRie87}. Recently higher-loop contributions to the
`classical' action have been investigated~\cite{MarSan99, KraWul99}.

\section*{Acknowledgments} The authors would like to thank 
K.~Chadan, R.~Dick, H.~Grosse, P.~Kulish, D.~Maison, G.~Meissner,
O.~P\`ene, P.~Pre\v{s}najder, T.~Sch\"ucker, S.~Theisen and G.~Velo
for enlightening conversations.  Two of them (RH) and (JM) would also
like to thank the Max-Planck-Institut f\"ur Physik in M\"unchen for
financial support and J. Wess for his hospitality there. The work was
partially supported by the DAAD under the PROCOPE grant number
PKZ~9822848.  The work of one of the authors (SC) was supported by the
Korean Ministry of Education (Project No. 1998-015-D00074).

\end{document}